\documentclass[12pt,journal, onecolumn]{IEEEtran}

\usepackage{cite}
\usepackage{amsfonts}
\usepackage{amsmath}
\usepackage{amsthm}
\usepackage{amssymb}
\usepackage{graphicx}
\usepackage{subfigure} 
\usepackage{textcomp}
\usepackage{latexsym}
\usepackage{hyperref}
\usepackage{float}
\usepackage{stfloats}
\usepackage{multirow}
\usepackage{adjustbox}
\usepackage{balance}
\usepackage{makecell}
\usepackage[table]{xcolor}
\usepackage{balance}
\usepackage[ruled]{algorithm2e}

\newcommand{\nop}[1]{}
\usepackage{xcolor}

\newtheorem{remark}{Remark}

\def\BibTeX{{\rm B\kern-.05em{\sc i\kern-.025em b}\kern-.08em
    T\kern-.1667em\lower.7ex\hbox{E}\kern-.125emX}}
\begin{document}


\title{BESS Aided Reconfigurable Energy Supply using Deep Reinforcement Learning for 5G and Beyond}

\author{Hao~Yuan,
		Guoming~Tang,
        Deke~Guo,
        Kui~Wu,
		Xun~Shao,
		Keping~Yu,
		Wei Wei

\thanks{H. Yuan and D. Guo are with the Science and Technology on Information Systems Engineering Laboratory, National University of Defense Technology, Changsha, Hunan, China. G. Tang is with the Peng Cheng Laboratory, Shenzhen, Guangdong, China. K. Wu is with the Department of Computer Science, University of Victoria, Victoria, BC, Canada. X. Shao is with the School of Regional Innovation and Social Design Engineering, Kitami Institute of Technology, Kitami, Japan. Keping Yu is with the Global Information and Telecommunication Institute, Waseda University, Shinjuku, Tokyo, Japan. W. Wei is wit School of Computer Science and Engineering, Xi'an University of Technology, Xi'an, China.}

\thanks{Corresponding authors: G. Tang and D. Guo.}
}

\maketitle\renewcommand{\thefootnote}{\arabic{footnote}}

\nop{
\author{Hao~Yuan\IEEEauthorrefmark{1},
       Guoming~Tang\IEEEauthorrefmark{1},
        Deke~Guo,
        Xueshan~Luo
\IEEEcompsocitemizethanks{\IEEEcompsocthanksitem
\IEEEauthorrefmark{1}Science and Technology on Information Systems Engineering Laboratory,
National University of Defense Technology, Changsha {\rm410073}, P.R. China.\protect
\IEEEauthorrefmark{2}Peng Cheng Laboratory, Shenzhen {\rm 518000}, P.R. China.\protect
\IEEEauthorrefmark{3}College of Intelligence and Computing, Tianjin University, Tianjin {\rm 300350}, P.R. China.
E-mail: yuanhao@nudt.edu.cn, tanggm@pcl.ac.cn, guodeke@gmail.com, xsluo@nudt.edu.cn. Corresponding authors: G. Tang and D. Guo.
}}
\maketitle\renewcommand{\thefootnote}{\arabic{footnote}}

\markboth{Journal of \LaTeX\ Class Files,~Vol.~?, No.~?, ?~20??}%
{Shell \MakeLowercase{\textit{et al.}}: Bare Demo of IEEEtran.cls for Computer Society Journals}
}

\maketitle

\begin{abstract}
The year of 2020 has witnessed the unprecedented development of 5G networks, along with the widespread deployment of 5G base stations (BSs). Nevertheless, the enormous energy consumption of BSs and the incurred huge energy cost have become significant concerns for the mobile operators. As the continuous decline of the renewable energy cost, equipping the power-hungry BSs with renewable energy generators could be a sustainable solution. In this work, we propose an energy storage aided reconfigurable renewable energy supply solution for the BS, which could supply clean energy to the BS and store surplus energy for backup usage. Specifically, to flexibly reconfigure the battery's discharging/charging operations, we propose a deep reinforcement learning based reconfiguring policy, which can adapt to the dynamical renewable energy generations as well as the varying power demands. Our experiments using the real-world data on renewable energy generations and power demands demonstrate that, our reconfigurable power supply solution can achieve an energy saving ratio of $74.8\%$, compared to the case with traditional power grid supply.
\nop{
The year of 2020 has witnessed the unprecedented development of 5G networks, along with the widespread deployment of 5G base stations (BSs). Compared to the 4G/LTE, 5G is supposed to provide much higher bandwidth, lower and more reliable latency, and larger number of connections for the massive IoT devices. Nevertheless, the enormous energy consumption of BSs and the incurred huge energy cost have become significant concerns for the mobile operators. As the continuous decline of the renewable energy price, equipping the power-hungry BSs with renewable energy generators could be a promising solution for energy cost reduction. In this work, we propose a battery energy storage system (BESS) aided renewable energy supply solution for the BS, which could supply clean energy to the BS and store surplus energy for backup usage. Specifically, to better control the battery's discharging/charging, we propose a deep reinforcement learning (DRL) based storage controlling policy, which can adapt to the dynamical renewable energy generations as well as the varying power demands. Our experiments using the real-world data on renewable energy generations and the power demands demonstrate that, the proposed solution can help save the monthly saving of one BS by up to \$50.7 (with a corresponding saving ratio of 74.8\%), compared to the case with only power grid supply.
}
\end{abstract}


\section{Introduction}
\begin{figure}[t!]
    \centering
	\centerline{\includegraphics[width=0.8\linewidth]{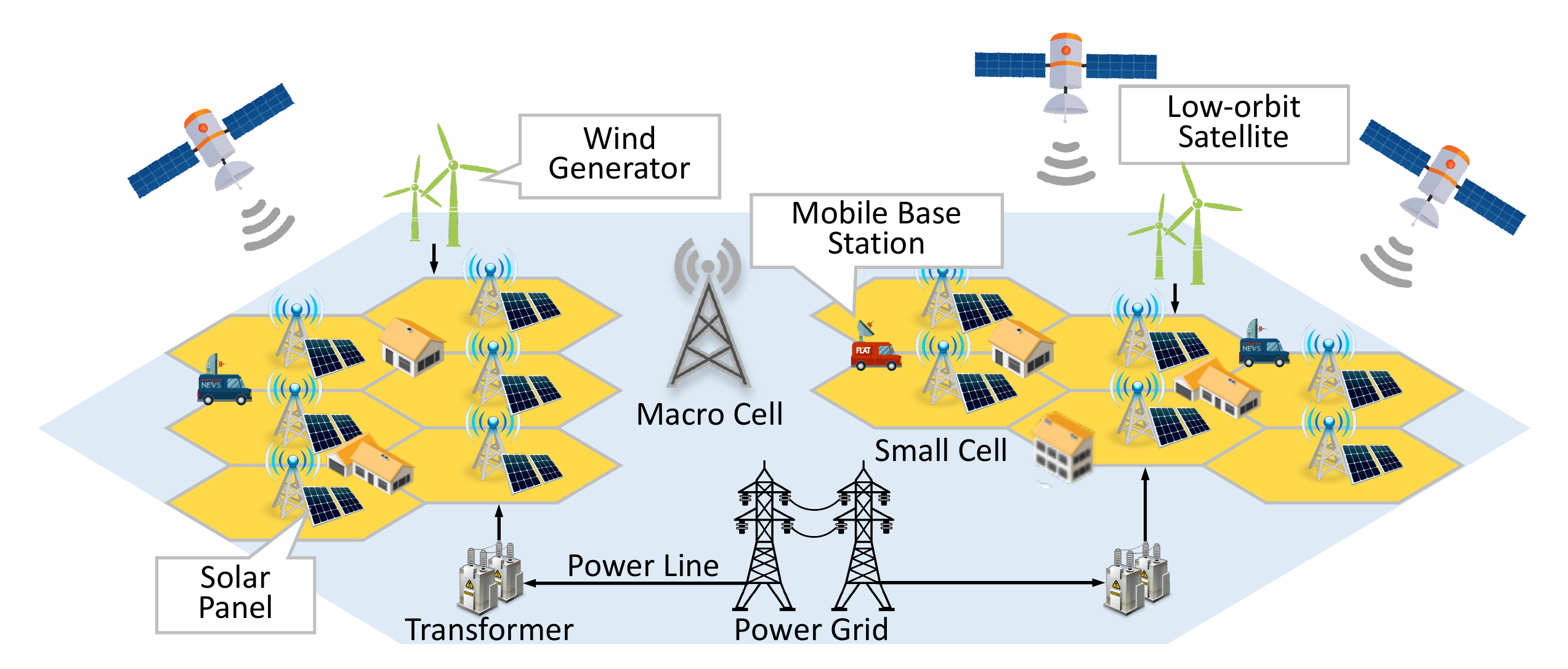}}
	\caption{A vision of the future radio access network (RAN) in 5G and beyond, which consists of macro and small cells, and also includes the mobile and space BSs. For the purpose of green communication, all the BSs could be supplied by both the renewable energy and power grid.}\label{fig:scenario}
	\vspace{-0.15in}
\end{figure}

The 5G network is considered as a promising technology to significantly improve the way how we live~\cite{andrews2014will}. Compared to the 4G/LTE, it can ensure users with higher bandwidth and lower latency and thus enable various cutting-edge mobile services, such as the Internet of Vehicles~\cite{gerla2014internet,kumar2015coalition}, Virtual Reality~\cite{burdea2003virtual}, and Smart Medical Home~\cite{muse2017towards,misra2020mac}. Nevertheless, due to the adoption of high frequency bands by 5G base station (BS), its signal coverage range is much shorter than that of the 4G/LTE. Consequently, the mobile operators need to deploy a large number of 5G BSs to tackle the problem of poor signal coverage. This would result in an ultra-dense BS deployment, especially in ``hotspot'' areas, as illustrated in Fig.~\ref{fig:scenario}. \nop{According to~\cite{huawei5g}, by 2026 and in China only, over 14 million 5G BSs will be deployed.}

Building and operating such large-scale BSs require an enormous investment and consume many resources (e.g., power consumption). According to field surveys in the cities of Guangzhou and Shenzhen, China, the full-load power consumption of a typical 5G BS is about $2\sim3$ times of that of a 4G BS~\cite{tang2020shiftguard}. Considering the ultra-dense deployment of 5G BSs, it could lead to a tenfold increase in energy consumption. In this regard, how to effectively reduce energy consumption becomes an urgent problem to be solved.

Renewable energies like the solar energy and wind energy, as the eco-friendly way of power supply with low $CO_2$ emissions, have been popularized in more scenarios in recent years. \nop{Compared with the cost of power generation with fossil fuels,} Owing to the continuing price decline in photovoltaic (PV) module and wind turbine, the installation cost of renewable energy has dramatically decreased over the past decade, e.g., it reports a 61\% reduction of the solar equipment from 2010 to 2017~\cite{fu2017us}. Such cost reductions lead to a rapid payback period for the renewable energy investment, from a couple of years to several months~\cite{turner1999realizable}. The above observations indicate the great potential of renewable energy on the market of fossil fuel replacement and carbon emission reduction.

\nop{Actual calculations show a very rapid payback of renewable energy~\cite{turner1999realizable}. In particular, the energy payback for the current solar energy has been calculated to range from 3 to 4 years, and wind energy has an even faster payback of 3 to 4 months. It makes renewable energy generators have great potential in energy-saving and replacement of fossil fuel during their lifetime (25-30 years)~\cite{lund2007renewable}. }

It thus has inspired the mobile operators to utilize renewable energy as the auxiliary power supply to tackle the huge power demand at 5G BSs. In some developing countries, solar power has already been applied to supply the BSs, some of which occupies over $8\%$ of the total electricity usage~\cite{wang2012survey}. By installing the PV and wind turbine near the BSs, it shows that the maximum power from the solar and wind generators can reach up to 8.5kW and 6.0kW, respectively~\cite{wang2012survey}, which could remarkably cut down the communication energy supply from the traditional power gird.

To maximize the utilization of renewable energy, \emph{energy storage} can be strategically utilized such that the energy can be continuously provided, as the renewable (like solar or wind) energy is intermittent and unstable. Meanwhile, most BSs are equipped with backup batteries to safeguard the BS's normal functioning against power outages, making it the natural energy storage. Besides, with the continuous price decline in battery storage these years~\cite{nykvist2015rapidly,mondal2015distributed}, combining the battery storage with renewable energy generators could offer even greater cost-reduction potential. Specifically, i) when the generated renewable power is less than the power demand (e.g., during the peak hours), the battery can be discharged to flatten the peak power demands, and ii) when the generated renewable power is more than the power demand (e.g., during the off-peak hours), the battery can be charged to store the surplus renewable energy.

In this paper, we propose a battery energy storage system (BESS) aided renewable energy supply solution for the 5G network and beyond. Aiming at energy cost reduction for mobile operators, our solution is to maximize the utilization of the renewable energy and thus minimize the utilization of power grid (i.e., fossil energy). Specifically, the energy charge can be continuously reduced by the generated renewable power, and the demand charge can be reshaped and flatten through strategic battery discharging/charging operations. \nop{Therefore, an effective design of a battery control strategy could significantly alleviate the increasing power demand. }

When designing the optimal control strategy in battery discharging/charging operations, we are faced with several challenges. Firstly, the renewable energy generation and power demand are highly varying in both spatial and temporal dimensions and thus hard to predict. Secondly, owing to the physical constraints of the battery discharging/charging operations (e.g., discharge/charge efficiency), it is complicated to design the optimal battery controlling policy. Thirdly, as the battery's capacity and lifetime are limited and shortened along with the discharge/charge cycles, it is necessary while non-trivial to trade-off between the cost of battery's degradation/replacement and the gain of renewable energy storage.

By tackling the above challenges, we make the following contributions in this work:
\begin{itemize}
    \item We present the BESS aided reconfigurable renewable energy supply paradigm for 5G BS operations, in which the battery discharging/charging reconfiguration is modelled as an optimization problem. The model is comprehensive by taking into account the practical considerations of dynamic power demand and renewable energy generation, as well as battery specifications and physical constraints.
    \item To cope with the intermittent renewable energy generation and dynamic BS power demand, while keeping computation complexity of the optimization problem under control, we propose a deep reinforcement learning (DRL) based battery discharging/charging reconfiguring policy, which can improve its decision-making efficiency through interacting with the environment.
    \item We conduct extensive evaluations using real-world BS deployment scenario and BS traffic load traces. The results show that the proposed DRL-based battery discharging/charging reconfiguring policy can effectively utilize the renewable energy and cut down the energy cost\nop{, with the cost saving up to $\$50.7$}.
\end{itemize}
\nop{
\begin{itemize}
    \item We present the BESS aided renewable energy supply paradigm for 5G BS operations, in which the battery discharging/charging controlling is modelled as an optimization problem. The model is comprehensive by taking into account the practical considerations of dynamic power demand and renewable energy generation, as well as battery specifications and physical constraints.
    \item To cope with the intermittent renewable energy generation and dynamic BS power demand, while keeping computation complexity of the optimization problem under control, we propose a deep reinforcement learning (DRL) based battery discharging/charging controlling policy, which can improve its decision-making efficiency through interacting with the environment.
    \item We conduct extensive evaluations using real-world BS deployment scenario and BS traffic load traces. The results show that the proposed DRL-based battery discharging/charging controlling policy can effectively utilize the renewable energy and cut down the energy cost\nop{, with the cost saving up to $\$50.7$}.
\end{itemize}
}

The rest of the paper is organized as follows. In Sec.~\ref{sec:background}, we introduce the background of this paper. In Sec.~\ref{sec:systemmodel}, we give the system models and formulations of the problem, and then propose the BESS aided renewable energy supply solution in Sec.~\ref{sec:power}. We develop a DRL-based battery discharging/charging controlling policy in Sec.~\ref{sec:solution}. We evaluate the proposed method by experiments with a real-world dataset in Sec.~\ref{sec:performance}. We present the related work in Sec.~\ref{sec:relatedwork} and conclude the paper in Sec.~\ref{sec:conclusion}.

\section{Background}\label{sec:background}

\begin{figure*}[!t]
 	\centering
 	 	\subfigure[Power demand pattern of BSs at resident area.]{
 		\includegraphics[width=0.32\textwidth]{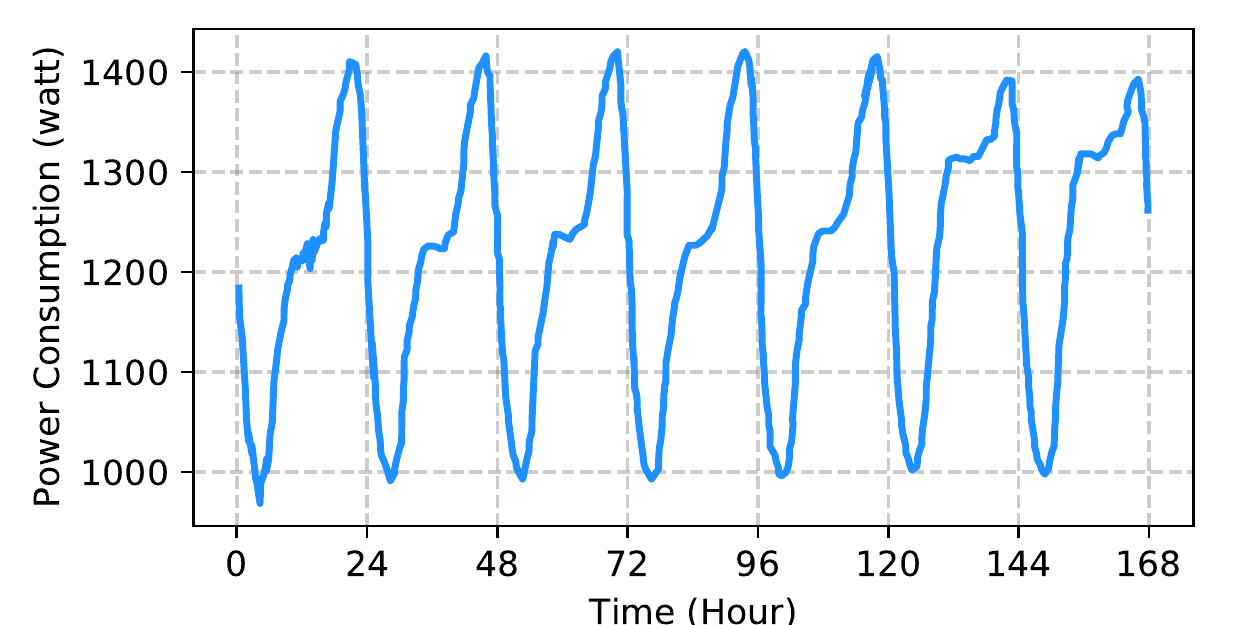}}
 	\hfill
 	\subfigure[Power demand pattern of BSs at office area.]{
 		\includegraphics[width=0.32\textwidth]{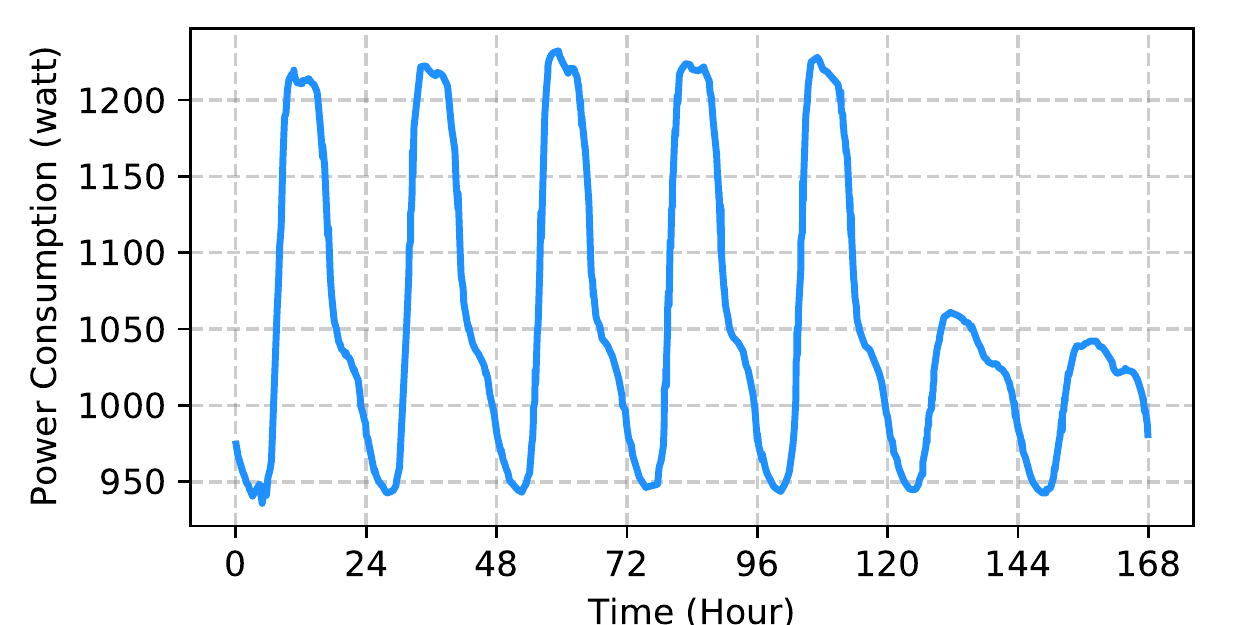}}
 	\hfill
 	\subfigure[Power demand pattern of BSs at comprehensive area.]{
 		\includegraphics[width=0.32\textwidth]{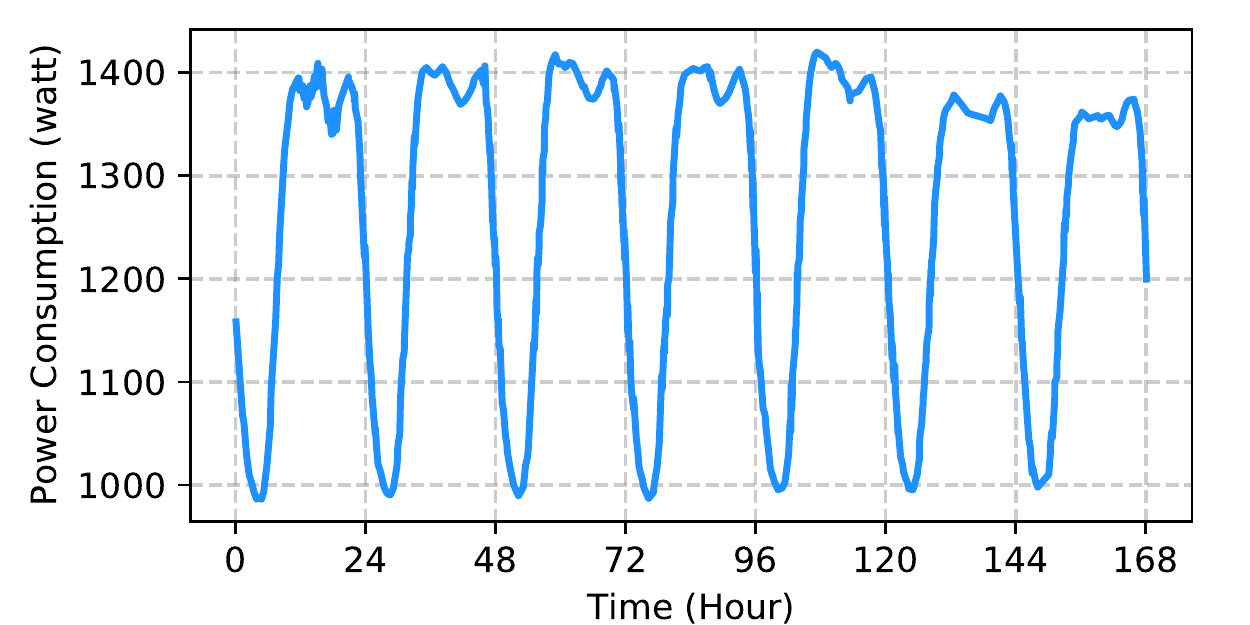}}
 	\caption{Power demand patterns of BSs at different area in one week period~\cite{wang2015understanding}.}
 	\label{fig:demand-pattern}
 	\vspace{-0.15in}
\end{figure*}

\subsection{Base Station Power Demand}
The power demand pattern of a BS is mainly determined by its location and associated with the behavior of users there. Usually the demand also show a periodic pattern (e.g., with a one-day or one-week period). As shown in Fig.~\ref{fig:demand-pattern}, in this paper, we mainly consider three types of BSs at the areas of \textit{resident}, \textit{office}, and \textit{comprehensive}, which account for nearly ninety percentage of the total demands~\cite{wang2015understanding}. To be detailed, the characteristics of these power demand patterns are as follows.
\begin{itemize}
    \item \textit{Power Demand of BSs at Resident Area}: The power demands of this type of BSs increase rapidly in the evening, as most people stay at home after work. Compared with those in weekdays, the power demands keep at high-levels in weekends.
    \item \textit{Power Demand of BSs at Office Area}: The power demands of this type of BSs keep at the high-level in the day time, when most people work during the time. Besides, due to the fewer people work on the weekends, the weekend power demands are much lower than those on the weekdays. 
    \item \textit{Power Demand of BSs at Comprehensive Area}: Due to the diversity of the requests, compared to the above two BSs, the power demand patterns of this type of BSs are more stable: constantly keep at a high-level in the day time and evening and drop down to the valley in late night and early morning.
\end{itemize}

The first two types of power demand patterns change relatively dramatically, leading to a huge energy-saving potential, especially for the \emph{demand charge}, which will be discussed in the next section. 

\subsection{Energy Cost of 5G BS}
The energy cost of the mobile operator typically makes up of two components: i) \textit{energy charge}, i.e., the total consumed electricity amount (in kWh) throughout the entire billing cycle (e.g., one month), and ii) \textit{demand charge}, i.e., the peak power demand (in kW) during the billing cycle period. Specifically, the demand charge is regarded as a penalty due to the caused extra load burden to the power grid. 

For example, for a commercial data center consuming 10 MW on peak and 6 MW on average, the monthly energy charge and demand charge amounts to around \$24,000 and \$165,500, respectively~\cite{xu2014reducing}. The demand charge could be up to 8x the energy charge, therefore, effectively cutting down the demand charge could remarkably reduce the energy cost. However, there seems no practical way to flatten the peak power demands of 5G BSs, e.g., shifting the real-time demands from mobile users to the off-peak hours could lead to the long delay for some of the classes of jobs~\cite{dabbagh2017shaving}.

\section{System Model}\label{sec:systemmodel}
In this section, we present the system models and basic assumptions and problem formulation. For clarity, the major notations used in this paper are explained in Table~\ref{tbl:notations}. 

\subsection{Scenario Overview}
\begin{figure}[t!]
    \centering
	\centerline{\includegraphics[width=0.8\linewidth]{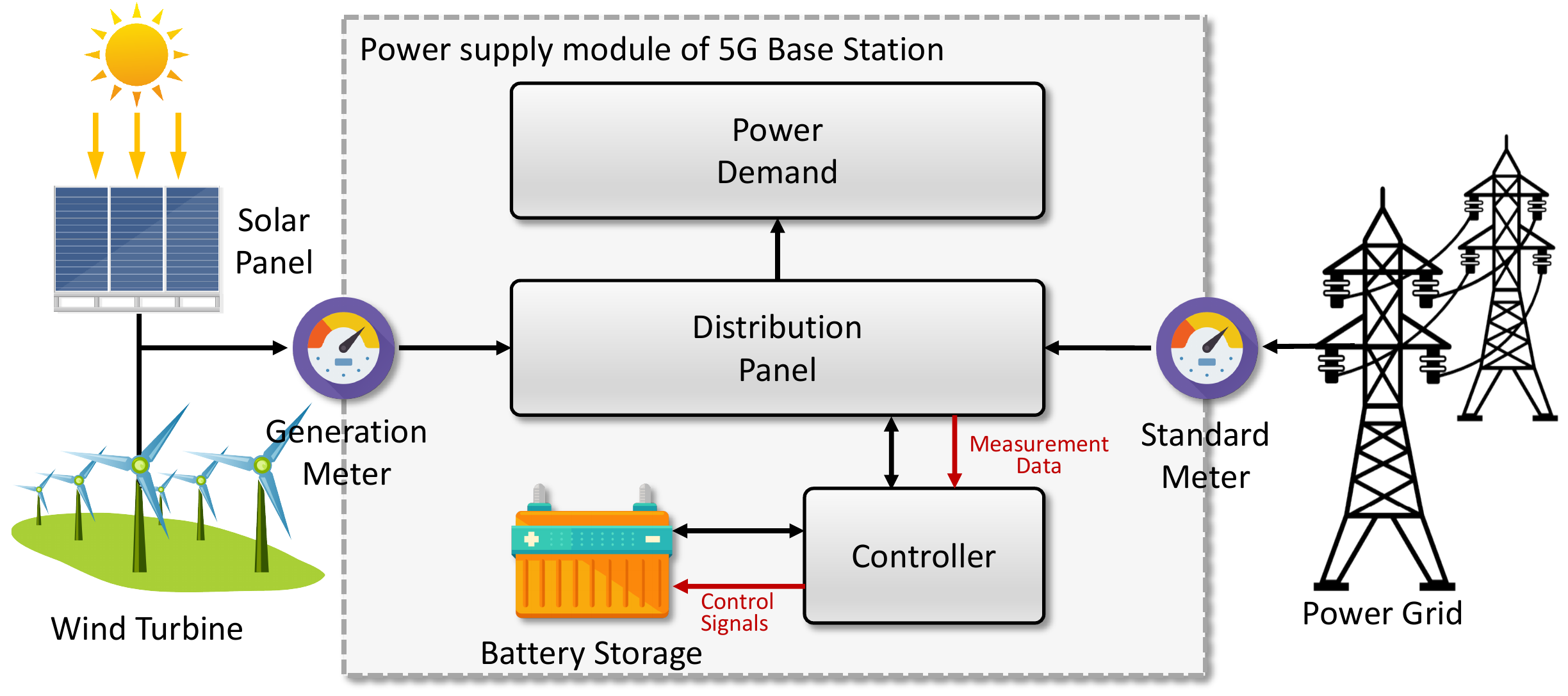}}
	\caption{An exemplified implementation of the BESS aided renewable energy supply solution for the 5G BS.}\label{fig:sys}
	\setlength{\abovecaptionskip}{0cm}
\end{figure}

As illustrated in Fig.~\ref{fig:sys}, the proposed BESS aided renewable energy supply solution deployed at each 5G BS mainly includes: i) a renewable energy generator, e.g., 
the PV panel and wind turbine, which is deployed near the 5G BS system and generates renewable energy for the system, ii) a battery storage, which stores the surplus renewable energy and acts as the power source for the BS as needed, and iii) a controller, which can obtain the environment state (i.e., the measurement data) so as to control the battery discharging/charging operations through the control signals. In addition to the standard meter, as shown in Fig.~\ref{fig:sys}, an additional generation meter is installed for the BS power supply system to measure the renewable energy generation. Furthermore, with commands from the controller, the distribution panel takes responsibility of power switch between the renewable energy and grid energy and ensures continuous and stable electricity supply for the BS.

As the essential component of the BESS aided renewable energy supply solution, the controller determines how efficient this paradigm is. Specifically, at each scheduling point, the controller needs to decide the amount of power supply from either the battery or the power grid. The scheduling operations should be made upon the power demands and battery states in real-time, so that the utilization of renewable energy can be enhanced and the total energy cost can be minimized.

Note that the feasibility of such an implementation as illustrated by Fig.~\ref{fig:sys} has been preliminarily verified in practice. According to~\cite{jinfan}, small integrated renewable energy generators are provided by some commercial companies for the BS system, which are easily deployed in both open rural and crowded urban environments.

\nop{
\begin{remark}
According to~\cite{jinfan}, some companies have already produced small integrated renewable energy generators at present, which can be easily deployed in both open rural and crowded urban environment. Therefore, our solution of deploying it near the 5G BSs is convenient and feasible.
\end{remark}
}

\subsection{BS Power Supply and Demand}
The power of each 5G BS is mainly supplied by three parts: power grid, generated renewable energy, and storage energy. In particular, i) when the generated renewable energy is more than the power demand (e.g., during the off-peak hours), each 5G BS is only supplied by the renewable energy (i.e., off-grid) and the surplus renewable energy is stored in the battery storage, ii) when the generated renewable energy is less than the power demand (e.g., during the peak hours), each 5G BS is supplied by all three parts in a cooperative way. 

In this paper, we consider a discrete time model, where the entire billing cycle (e.g., one month) is equally spilt into $T$ consecutive slots with length of $\Delta t$ and denoted by $\mathcal{T} = \{ 1, 2, \cdots,T \}$. For an arbitrary 5G BS, the power demand during the entire billing cycle can be represented by a power demand vector:
\begin{equation}
    d := [d(1), d(2), \cdots, d(T)]
\end{equation}
where $d(t)$ is the power demand in time slot $t$, which can be obtained by power meter readings at each BS.

\subsection{Renewable Energy Generation}\label{sec:renewable}
By harvesting energy from renewable energy resources, the BSs could be powered in an environmentally-friendly and cost-efficient way. In this paper, in order to make the model extensible, we denote the renewable energy generation vector as:
\begin{equation}
    g := [g(1), g(2), \cdots, g(T)]
\end{equation}

In this work, we choose two typical renewable energy as the auxiliary way of power supply, i.e., solar energy (i.e., $g^s(t)$) and wind energy (i.e., $g^w(t)$). Accordingly, for an arbitrary time slot $t$, the renewable energy generation vector can be represented by:
\begin{equation}
    g(t) = g^s(t) + g^w(t)
\end{equation}

We assume that if the total generated renewable energy is beyond the power demand (i.e., $g(t) > d(t)$), the power is supplied in proportion to the renewable energy generated. The generation of both varies during a certain period (e.g., one day) and is affected by a some similar factors such as weather, temperature, wind speed, and so on.

\subsubsection{Solar Energy Generation}
Power generated by the solar PV system mainly depends on three factors: global horizontal irradiance ($GHI(t)$), outdoor temperature ($Temp(t)$), and time of day ($ToD(t)$). By arranging solar PV cells in series/parallel, solar PV could harvest energy and convert it into DC to charge the battery storage and supply the power demand. The generated power by the solar PV at time slot $t$ can be measured by the following function:
\begin{equation}
    g^s(t) = \mathbb{F}^S ( GHI(t), Temp(t), ToD(t))
\end{equation}
where $\mathbb{F}^S(\cdot)$ is a known, non-linear function defined in PVLIB~\cite{holmgren2015pvlib}. Accordingly, the solar energy generation during the entire billing cycle can be represented by a vector:
\begin{equation}
    g^s := [g^s(1), g^s(2), \cdots, g^s(T)]
\end{equation}

\subsubsection{Wind Energy Generation}
Power generated by the wind turbine generator fluctuates randomly with time and mainly depends on the wind velocity ($WV(t)$), weather system ($WS(t)$), and hub height ($HH(t)$). The wind turbine generate energy typically into two stages: first, it converts the wind power into mechanical energy and then transforms into electricity. The amount of the power generated by the wind turbine at time slot $t$ can be calculated by the following function:
\begin{equation}
    g^w(t) = \mathbb{F}^W ( WV(t), WS(t), HH(t))
\end{equation}
where $\mathbb{F}^W(\cdot)$ is a known, non-linear function defined in~\cite{jahid2020techno}. Accordingly, the wind energy generation during the entire billing cycle can be represented by a vector:
\begin{equation}
    g^w := [g^w(1), g^w(2), \cdots, g^w(T)]
\end{equation}

\subsection{Battery Specification}
At an arbitrary time slot $t$, the state of the battery is modeled as follows:
\begin{equation}
    \chi(t) := \langle SoE(t), SoC(t), DoD(t) \rangle
\end{equation}
where the notations of SoE, SoC, and DoD represent the \textit{state of effective capacity} \textit{state of charge}, and \textit{depth of discharge} of the battery, respectively. Specifically, i) \textbf{SoE} indicates the current effective capacity of the battery, as a percentage of its initial capacity (denoted as $\pi$), ii) \textbf{SoC} indicates the current energy stored in the battery, as a percentage of the current effective capacity, and iii) \textbf{DoD} indicates how much energy the battery has released, as a percentage of the current effective capacity.

For simplicity to tackling the optimization problem, we discretize the SoC of a battery into $M$ equal-spaced states (e.g., $M = 10$, i.e., $\{10\%, 20\%, \cdots, 100\% \}$). Accordingly, the DoD are also discretized (e.g., release $10\%$ from $90\%$, i.e., $90\% \to 80\%$). Besides, for an arbitrary time slot $t$, in order to prevent the battery from over-discharging/charging, we use $SoC_{max}$ and $SoC_{min}$ to indicate the upper and lower bounds of SoCs, respectively, which is shown as follows.
\begin{equation}\label{eq:8}
    SoC_{min} \leq {SoC}(t) \leq SoC_{max}
\end{equation}

\renewcommand{\arraystretch}{1.1}
\begin{table}[!t]
    \centering
	\caption{Summary of notations}\label{tbl:notations}
	\begin{tabular}{r l}
		\Xhline{2\arrayrulewidth}
		\emph{Notation}        & \emph{Description}  \\ 
		\hline
    	$d(t)$        		&power demand of 5G BS in time slot $t$ \\
    	$g(t)$             &renewable energy generation in time slot $t$ \\
    	$b(t)$               &battery discharging/charging operations in time slot $t$ \\
    	$\chi(t)$            &battery state in time slot $t$ \\
    	$p(t)$               &power supplied by the power gird in time slot $t$ \\
        $p_{max}$            &peak power consumption supplied by power gird \\
        $\pi$                &initial capacity of the battery\\
		\hline
		$\mathcal{C}^e(t)$      &energy charge of 5G BS in time slot $t$ \\
		$\mathcal{C}^d(t)$      &demand charge of 5G BS in time slot $t$ \\
		$\mathcal{C}^u(t)$      &investment cost in time slot $t$ \\
		$\lambda_{e}$       & prices of energy charge\\
		$\lambda_{d}$       & prices of demand charge\\
		$\lambda_{u}$       & prices of investment cost\\
		$\alpha, \beta$		& discharging and charging efficiencies, respectively \\
		$R+, R-$			& max charge and discharge rates of battery, respectively \\
		\hline
		$s(t)$		& environment state in time slot $t$ \\
		$a(t)$			& action taken by the agent in time slot $t$ \\
		$r(t)$          & reward of the action in time slot $t$ \\
		$\psi$		& mapping policy from environment states to actions \\
    	$R(a(t),s(t))$			  & reward function of the DQN \\
    	$Q, \tilde{Q}$	  & Q-values of the main net and target net, respectively \\
    	$\theta, \tilde{\theta}$ & parameters of the main net and target net, respectively \\
    	
    	\Xhline{2\arrayrulewidth}
	\end{tabular}
	\vspace{-0in}
\end{table}

\section{BESS Aided Renewable Energy Supply}\label{sec:power}
The battery storage is deployed at 5G BS, and can charge by the surplus renewable energy (generated by solar PV and wind turbine system) and discharge to reshape the power demand, so as to maximize the utilization of renewable energy (or minimize the utilization of fossil fuel) and reduce the electricity bill.

We define the battery discharging/charging operations by a \textit{battery operation vector}:
\begin{equation}
    {b} := [{b}(1), {b}(2), \cdots, {b}(T)]
\end{equation}
where $b(t)$ is a real number variable and indicates the amount of discharging/charging operations. To be detail, i) \textbf{positive value} indicates discharging the power from the battery storage to the 5G BS during time slot $t$, ii) \textbf{negative value} indicates charging from the renewable energy to the battery storage, and iii) \textbf{zero value} indicates no discharging/charging operation performs. 

Meanwhile, the discharging/charging operations is constrained by the maximum charging rate and maximum discharging rate, denoted as $R^+$ and $R^-$, respectively. It means the the largest power that the battery can be recharged and supply with in a time slot, which is shown as follows. 
\begin{equation}\label{con-21}
-R^{+} \leq b_{n}(t) \leq R^{-}
\end{equation}

Besides, the battery storage need to meet the following conditions in discharging/charging operations:
\begin{subequations}\label{eq:11}
\begin{align}
{b}(t) \le 0 & \mbox{, if } g(t)-d(t) \ge 0\\
{b}(t) > 0 & \mbox{, if } g(t)-d(t) < 0
\end{align}
\end{subequations}
which represents that the battery storage can only be charged when there exists surplus renewable energy after supplying to the 5G BS, and means that the battery storage cannot be simultaneously charged and discharged at any time slot. 

Due to the power loss (e.g., AC-DC conversion and battery leakage~\cite{qi2019energyboost}) occurred during discharging from battery storage to the power grid (or charging from renewable energy to the battery storage), we denote the actual discharging/charging operations from/to the battery by:
\begin{equation}
\tilde{b}(t) = \left\{\begin{array}{cl}
b(t)/\alpha & \mbox{, if } {b}(t) \leq 0 \\
\beta \cdot {b}(t) & \mbox{, if } b(t) > 0 
\end{array}\right.
\end{equation}

Given the power demand of the 5G BS (i.e., $d(t)$), the renewable energy generation (i.e., $g(t)$), and the battery discharging/charging operations (i.e., $b(t)$), we can derive the
\textit{power consumption vector} supplied by the power grid for an arbitrary time slot $t$ by:
\begin{equation}
    p := [p(1), p(2), \cdots, p(T)]
\end{equation}
where $p(t)$ is denoted as:
\begin{equation}
p(t) = \left\{\begin{array}{cl}
max \{0, d(t)-g(t)-\tilde{b}(t) \} & \mbox{, if discharging}\\
max \{0, d(t)-g(t) \} & \mbox{, if charging}
\end{array}\right.
\end{equation}

\subsection{Energy Cost}
The billing policy of the energy cost for the mobile operators throughout the entire billing cycle typically make up of two components, energy charge and demand charge, which is widely applied in previous~\cite{xu2014reducing, dabbagh2017shaving, shi2016leveraging}. And we will introduce them in detail as follows.

\begin{itemize}
    \item Energy Charge: the total consumed electricity amount (in kWh) throughout the entire billing cycle (in the unit \$kWh and denoted by $\lambda_e$).
    \item Demand Charge: the peak power consumption supplied by power gird (in kW) during the entire billing cycle (in the unit \$kW and denoted by $\lambda_d$).
\end{itemize}

Therefore, the incurred cost of \textbf{energy charge} of the whole system in each time slot $t$ can be represented by:
\begin{equation}
    \mathcal{C}^e(t) = \lambda_e \cdot p(t) \cdot \Delta t
\end{equation}
Accordingly, the incurred cost of \textbf{demand charge} of the whole system in each time slot $t$ can be represented by:
\begin{equation}
    \mathcal{C}^d(t) = \lambda_d \cdot max \big\{ 0, p(t) - p_{max} \big\}
\end{equation}
where $p_{max}$ records the peak power consumption during the past $t-1$ time slots. For any arbitrary time slot $t$, if $p(t) - p_{max} > 0$, $p_{max}$ will be updated to $p(t)$ accordingly.

\subsection{Investment Cost}
Every usage of this equipment (solar PV, wind turbine, and battery storage) incurs a certain reduction of its lifetime, which is essential for the investor. Therefore, it is significant to understand, detail and quantify the various factors influencing the performance loss curves. For the accuracy of our model, we quantify the investment cost in every time slot as follows.

\subsubsection{Renewable Energy Generator Cost}
As modules of a renewable energy generated system age, they gradually lose some performance. In this paper, we assume the decline of the system is linear and positively related to its using time. We denote the lifetime of the renewable energy generator as $L$, which indicates the total time it can be used. For an arbitrary time slot $t$, the remaining lifetime of the renewable energy generator is denoted as $l(t)$, which is constrained by $0 \le l(t) \le L$. The renewable energy generator has to be discarded and replaced by a new one if $l(t) \le 0$. Given the remaining lifetime of the renewable energy generator at time $t-1$, the remaining lifetime at time $t$ is updated by:
\begin{equation}
    l(t) = l(t-1) - \Delta t \cdot u(t)
\end{equation}
where $u(t)$ is defined by:
\begin{equation}
u(t) = \left\{\begin{array}{cl}
1 & \mbox{, if using}\\
0 & \mbox{, if not using}
\end{array}\right.
\end{equation} 
We formulate the using cost of the renewable energy generator in each time slot $t$ as: 
\begin{equation}\label{eq:usingcost}
    \mathcal{C}^{u}(t) = \lambda \cdot \frac{\Delta t \cdot u(t) }{L}
\end{equation}
where $\lambda$ is the investment cost of a new renewable energy generator. 

We extend the model of renewable energy generator to specific system, i.e., the solar PV system and wind turbine system. To be detail, i) for the solar PV system, we denote the lifetime, the investment cost, and investment as $l^s(t)$, $\mathcal{C}^{u_s}(t)$, and $\lambda_{s}$, respectively, ii) for the wind turbine system, we denote the lifetime, the using cost, and investment as $l^w(t)$, $\mathcal{C}^{u_w}(t)$, and $\lambda_{w}$. Accordingly, we can derive the using cost of the solar PV system and wind turbine system by replacing the symbol in the Eq.~\ref{eq:usingcost}.

\subsubsection{Battery Storage Degradation Cost}
Every cycle of discharge/charge operation does some “harm” to the battery (typically lead-acid) and reduces its capacity and lifetime. Especially, a deep discharging severely affect its internal structure, even can permanently damage the battery (e.g., an overdischarging). The battery has to be discarded and replaced by a new one, when the effective capacity drops down to the "ineffective" level, denoted by $SoE_{ine}$ in this paper.

\begin{figure}[h]
    \centering
	\centerline{\includegraphics[width=9cm]{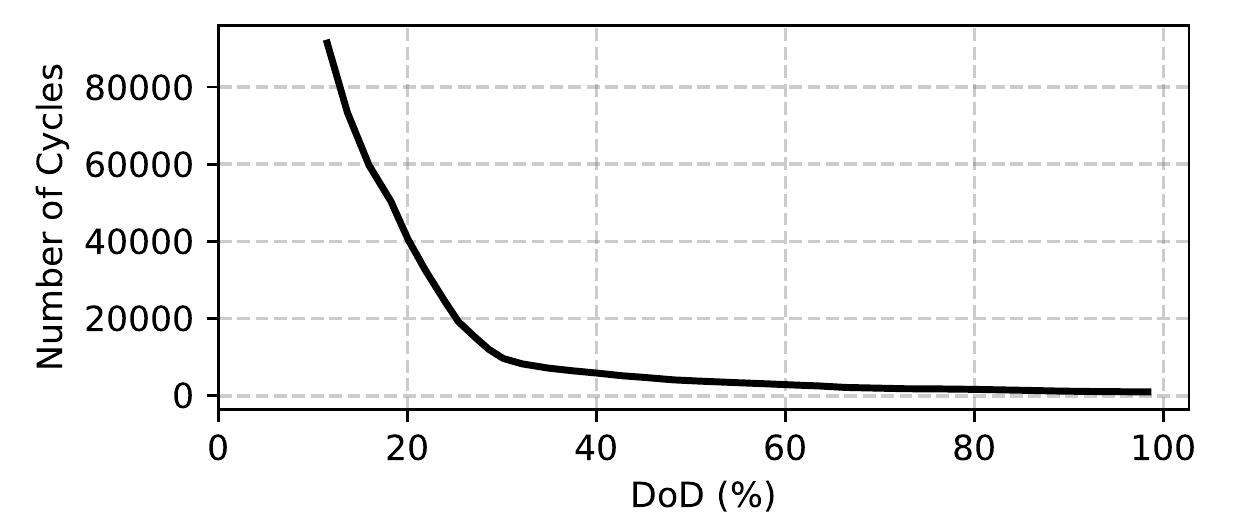}}
	\caption{Relationship between DoD levels and battery lifetime (in number of discharge/charge cycles) for LI battery, respectively~\cite{aksanli2013distributed}.}\label{fig:dod}
	\setlength{\abovecaptionskip}{0cm}
\end{figure}

As illustrated in Fig.~\ref{fig:dod}, each level of DoD has a corresponding number of discharge/charge cycles, thus, we can formulate the battery storage degradation cost by the relationship between both. Given a state of battery at time slot $t$, i.e., $\langle SoE(t), SoC(t), DoD(t) \rangle$, the SoE decrease of the battery during this time slot can be measured by:
\begin{equation}\label{dSoH}
\Delta{SoE}(t) = \left\{\begin{array}{cl}
0 & \mbox{, if } b(t) \leq 0 \\
\frac{1-SoE_{ine}}{h\left(DoD(t-1) +\Delta{DoD}(t)\right)} & \mbox{, if } b(t) > 0 
\end{array}\right.
\end{equation}
where $h(\cdot)$ maps from an input DoD level to the total number of discharge/charge cycles (exemplified in Fig.~\ref{fig:dod}), and $\Delta{DoD}(t)$ gives the increase of DoD and can be calculated by:
\begin{equation}\label{dDoD}
\Delta{DoD}(t) = \frac{b(t) \Delta{t}}{\pi}
\end{equation}

With the above expression of SoE decrease in each time slot $t$, we can then formulate the \textbf{degradation cost} of the battery storage at each time slot $t$ as:
\begin{equation} 
    \mathcal{C}^{u_b}(t) = \lambda_b \cdot \Delta{SoE}(t)
\end{equation}
 where $\lambda_{b}$ is a coefficient converting the battery degradation to a monetary cost, with the unit of “\$/SoE decrease”.

To sum up, the total \textbf{investment cost} in each time slot $t$ can be calculated as:
\begin{equation} 
    \mathcal{C}^{u}(t) = \mathcal{C}^{u_s}(t) + \mathcal{C}^{u_p}(t) + \mathcal{C}^{u_b}(t)
\end{equation}

\subsection{Optimization Formulation and Difficulty Analysis}
The battery discharging/charging operations is controlled by the controller. Given the state (i.e., $\chi (t)$) of the battery storage in time slot $t-1$, the state in time slot $t$ can be updated by:
\begin{equation}\label{con-23}
\chi(t) \leftarrow \left\{\begin{array}{lll}
SoE(t) & = & SoE(t-1) - \Delta SoE(t) \\
SoC(t) & = & SoC(t-1) - b(t)\Delta t/{\pi} \\
DoD(t) & = & DoD(t-1) + \Delta DoD(t)
\end{array}\right.
\end{equation}

For the entire billing cycle $\mathcal{T}$, we need to find the optimal battery discharging/charging controlling policy to solve the optimization problem, so as to minimize the total electricity bill during the entire billing cycle, which is defined as follows.
\begin{subequations}\label{opt-2}
\begin{align}
& \underset{b(t)}{\text{min}} && \sum_{t=1}^T \big ( \mathcal{C}^{e}(t) + \mathcal{C}^{d}(t) + \mathcal{C}^{u}(t) \big ) \\ 
& \text{s.t.} && (\ref{eq:8}), (\ref{con-21}), (\ref{eq:11} ), \text{and } (\ref{con-23}), \forall t \in \mathcal{T}
\end{align}
\end{subequations}

When solving the above optimization problems, however, we are faced with the following three challenges.

\subsubsection{Uncertainty of Renewable Energy}
Renewable energy generation is affected by multiple factors such as outdoor temperature and wind velocity. It is hard to accurately forecast renewable energy generation (i.e., $g(t)$) and make the optimal discharging/charging operations (i.e., $b(t)$) of the battery storage without accurate information in advance, as the unpredictable and intermittent nature of these factors. Therefore, we need to propose a method to tackle the problem of the uncertainty of renewable energy generation.

\subsubsection{Dynamic of Power Demand}
In our modeled problem, we assume the power demand (i.e., $p(t)$) is known in advance and thus can essentially optimize in an offline way. However, such assumptions are unrealistic in practice. In fact, traditional offline optimization methods (e.g., dynamic programming\cite{maly1995optimal, oudalov2007sizing}) are hard to find the global optimal solution, as the power demand can be obtained only when the workload arrives at the 5G BS. Thus, an online method to deal with the dynamic power demands (i.e., $d(t)$), and make optimal discharging/charging operations (i.e., $b(t)$), is in great need.

\subsubsection{High Computation Complexity} 
The optimization problem in Eq.~\ref{opt-2} has embedded NP-hard subproblems. Firstly, in every time slot $t$, the controller needs to search the action space (mainly determined by $M$), so as to find the the optimal discharging/charging operation (i.e., $b(t)$). For simplicity to solving the optimization problem, in this paper, we discretize the SoC of battery in to $M$ equal-spaced states, however, in real scenario, the state of the battery is continous, which leads to an enormous searching space. Secondly, during the entire billing cycle (i.e., $\mathcal{T}$), it is challenging for the controller to continuously make the optimal discharging/charging operation.

To tackle the above three challenges, we propose an online discharging/charging operation controlling method based on deep reinforcement learning (DRL) in the following section.

\section{A DRL-based Battery Operation Approach}\label{sec:solution}
Recent breakthrough of deep reinforcement learning (DRL)~\cite{mnih2015human} provides a promising technique for enabling effective experience-driven control, which exploit the past experience (e.g., historical battery discharging/charging operations) for better decision-making by adapting to current state of environment. We consider DRL is particularly suitable for online discharging/charging operation controlling because: i), it is capable of handling a high-dimensional state space (such as AlphaGo~\cite{silver2016mastering}), which is more advantageous over traditional Reinforcement Learning (RL)~\cite{sutton2018reinforcement}, and ii) it is able to deal with highly dynamic time-variant environments such as time-varying power demand and renewable energy generation. Next, we will introduce the basic components and concepts of DRL and the proposed DRL-based battery discharging/charging controlling policy in detail. 

\subsection{Components \& Concepts}
A typical DRL framework consists of five key components: \textit{agent}, \textit{state}, \textit{action}, \textit{policy}, \textit{and reward}. The concept and design of each component in our DRL-based battery discharging/charging controlling policy is explained as follows.

\begin{itemize}
    \item \textbf{Agent}: The role of the agent is to make decisions in every episode by interacting with the environment. Specifically, at the beginning of each time slot, it determines the discharging/charging operations (i.e., $b(t)$) according the current state (e.g., $d(t)$, $g(t)$ and $\chi (t)$) of the environment. The objective is to find an optimal battery discharging/charging controlling policy to minimize the total electricity bill during the entire billing cycle.
    
    \item \textbf{State}: At each episode, the agent first observes the state of the current environment to take action. In order to take the optimal action at each episode, the current state should cover as much information as possible. In this paper, we define the \textit{state vector} of the current environment as $s(t) = [d(t), g(t), \chi(t), p_{max}]$, which concludes the current information of the power demand, the renewable energy generation, the battery storage and the peak power consumption.
    
    \item \textbf{Action}: After observing the state of the environment, the agent will take an action accordingly. In our problem, the action is to control the battery discharging/charging operations in each time slot, i.e., $b(t)$, specifically, i) whether the battery should be discharged or charged, and ii) how much energy should be discharged or charged. We denote the action taken at time $t$ by $a(t)$, which is equivalent to $b(t)$.

    \item \textbf{Policy}: The battery discharging/charging controlling policy $\psi(s(t)): \mathcal{S} \to \mathcal{A}$ defines the mapping relationship from the state space to the action space, where $\mathcal{S}$ and $\mathcal{A}$ represent the state space and the action space, respectively. Specifically, the controlling policy can be represented by set of $a(t) = \psi(s(t))$, which maps the state of the environment to the action at time slot $t$.
    
    \item \textbf{Reward}: After interacting with the environment, the agent will receive a reward $r(t)$ (calculated by the reward function $R(s(t), a(t))$), which indicates the effect of the action in this episode, so as to update the controlling policy. The objective of the agent is to find a policy $\psi$ to maximize the total reward through continuous interacting with the environment. The design of the reward function significantly affect the performance of the DRL-based algorithm, and we will introduce its detail in the next subsection.
    
\end{itemize}

To sum up, at each episode, the agent observes the state $s(t)$, takes an action $a(t)$ generated by the policy $\psi$, and receives a reward $r(t)$ calculated by the reward function $R(s(t), a(t))$. The objective of the proposed DRL-based battery discharging/charging controlling policy is to take the optimal action in every episode so as to maximize the total reward.

\subsection{Reward Function Design}
At the end of each time slot, the agent evaluates the performance of the action using a reward function, which transforms the performance statistics to a numerical utility value. For an arbitrary time $t$, the agent observes the state $s(t)$, takes the action $a(t)$ and adopts the following reward function to access the performance of the controlling action:
\begin{equation}
    R(s(t), a(t)) = exp \big( V^e(t) + V^d(t) + V^u(t) \big)
\end{equation}
in which:
\begin{itemize}
    \item $V^e(t) = -\mathcal{C}^e(t)$, measures the reward of the incremental energy charge caused by the action in time slot $t$.
    \item $V^d(t) = -\mathcal{C}^d(t)$, measures the reward of the incremental demand charge caused by the action in time slot $t$.
    \item $V^u(t) = -\mathcal{C}^u(t)$, measures the reward of the investment cost caused by the action in time slot $t$.
\end{itemize}

At the end of each time slot, the agent evaluates the performance of the action by the reward $r(t)$ calculated by the reward function $R(s(t), a(t))$. In the DRL-based framework, the objective is to maximize the expected cumulative discounted reward:
\begin{equation}
r(t) = \mathbb{E} \big[ \sum_{k=t}^\infty {\gamma}^k R(s(t),a(t)) \big]
\end{equation}
where $\gamma \in (0,1]$ is a factor discounting future rewards. 

\begin{figure}[t!]
    \centering
	\centerline{\includegraphics[width=0.8\linewidth]{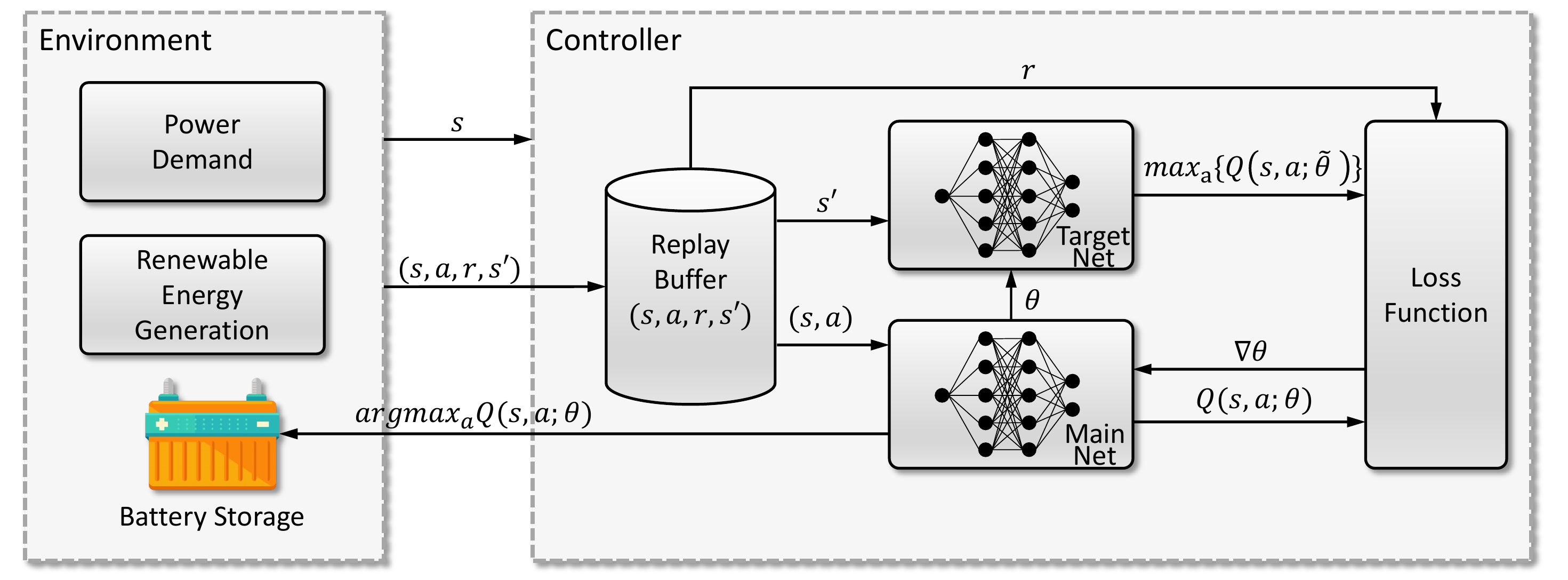}}
	\caption{The framework of the learning process in DQN. For simplicity, we denote $s(t+1)$ as $s'$. After interacting with the environment, the agent (i.e., controller) will determine the specific discharging/charging operation.}\label{fig:DQN}
	\setlength{\abovecaptionskip}{0cm}
\end{figure}

\subsection{Learning Process Design}
The learning process of the algorithm adopts a deep neural network (DNN) called Deep Q-Network (DQN) to derive the correlation between each state-action pair $(s(t), a(t))$ and its value function $Q(s(t), a(t))$, which is the expected discounted cumulative reward. If the environment is in state $s(t)$ and follows action $a(t)$, the value function of the state-action $(s(t), a(t))$ can be represented as:
\begin{equation}
Q(s(t), a(t)) = \mathbb{E} \big[ r(t) | s(t), a(t) \big]
\end{equation}

After obtaining the value of each state-action $(s(t), a(t))$, the agent selects the action $a(t)$ with the $\epsilon$-greedy policy $\psi$, that is, randomly selects the action with the probability of $\epsilon$, and chooses the action with the maximum of $Q(s(t), a(t))$ with the probability of 1-$\epsilon$, i.e., $\mbox{argmax}_{a(t)}Q(s(t), a(t))$. 

As illustrated in Fig.~\ref{fig:DQN}, two effective techniques were introduced in~\cite{mnih2015human} to improve stability: \textit{replay buffer} and \textit{target network}. Specifically,
\begin{itemize}
    \item \textbf{Replay Buffer}: Unlike traditional reinforcement learning, DQN applies a replay buffer to store state transition samples in the form of $\langle s(t), a(t), r(t),s(t+1) \rangle$ collected during learning. Every $\kappa$ time steps, the DRL-based agent updates the DNN with a mini-batch experiences from the replay buffer by means of stochastic gradient descent (SGD): $\theta_{i+1}=\theta_{i}+{\sigma}{\bigtriangledown}_{\theta}Loss(\theta)$, where $\sigma$ is the learning rate. Compared with Q-learning (only using immediately collected samples), randomly sampling from the replay buffer allows the DRL-based agent to break the correlation between sequentially generated samples, and learn from a more independently and identically distributed past experiences. Thus, the replay buffer can smooth out learning and avoid oscillations or divergence.
    \item \textbf{Target Network}: There are two neural networks with the same structure but different parameters in DQN, the main net and the target net. $Q(s,a;\theta)$ and $Q(s,a;\tilde{\theta})$ represent the current Q-value and target Q-value generated by the main net and the target net, respectively. The DRL-based agent uses the target net to estimate the target Q-value $\tilde{Q}$ for training the DQN. Every $\tau$ time steps, the target net copies the parameters from the main net, whose parameters are updated in real-time. After introducing the target net, the target Q-value will remain unchanged for a period time, which reduces the correlation between the current Q-value and the target Q-value and improves the stability of the algorithm.
\end{itemize}

Accordingly, the DQN can be trained by the loss:
\begin{equation}\label{eq:loss}
Loss(\theta)\leftarrow \mathbb{E} \big[(\tilde{Q}-Q(s(t),a(t);\theta))^2 \big]
\end{equation}
where $\theta$ is the network parameters of the main net, and $\tilde{Q}$ is the target Q-value and calculated by:
\begin{equation}\label{eq:loss_q}
\tilde{Q}\leftarrow r(t)+\gamma {max}_{a(t+1)}Q(s(t+1),a(t+1);\tilde{\theta})
\end{equation}
where $\tilde{\theta}$ is the network parameters of the target net and it updates every $\tau$ time slots by coping from the main net.

\begin{algorithm}[t!]
\small
\caption{Battery Controlling Algorithm with DRL}\label{alg:DRL}
\LinesNumbered
\KwIn{Power demand of BS $d(t)$ and renewable energy generation $g(t)$, $1 \le t \le T$}
\KwOut{Discharging/charging actions $a(t)$, $1 \le t \le T$}
Initialize replay buffer (RB) to capacity N\;
Initialize main net $Q$ with random weights $\theta$\;
Initialize target net $\tilde{Q}$ with weights $\tilde{\theta}=\theta$\;
\For {$episode = 1:MaxLoop$}{ 
    \For {$t=1:T$}{
        Get environment state $s(t)$ \;
        $a(t) = \left\{\begin{array}{l}
        \mbox{argmax}_{a}Q(s(t),a(t);\theta), \mbox{ prob. } \epsilon\\
        \mbox{random action,} \mbox{ prob. } 1-\epsilon
        \end{array}\right.$
        
        Execute action $a(t)$ and receive $r(t)$ and $s(t+1)$\;
        Store $\langle (s(t), a(t), r(t),s(t+1) \rangle$ into RB\;
        Randomly sample a mini-batch of experience $\langle s(i), a(i), r(i), s(i+1) \rangle$ from RB by every $\kappa$ steps\;
        $\tilde{Q} = \left\{\begin{array}{l}
            r(t), \mbox{ terminates at step } t+1 \\
            r(t)+{\gamma}\mbox{max}_{a(t+1)}\{{Q}(s(t+1),a(t+1); \tilde{\theta})\}, \text{ else}
        \end{array}\right.$
        
        Perform SGD on $(\tilde{Q}-Q(s, a; \theta))^2$ w.r.t. $\theta$\;
        Set $\tilde{Q}=Q$ by every $\tau$ steps\;
    }}
\end{algorithm}

To sum up, the learning process is depicted by the pseudo-code in Alg.~\ref{alg:DRL}. The controller first initializes the replay buffer and the parameters (i.e., $\theta$ and $\tilde{\theta}$) of the main net and target net, respectively. After obtaining the value of each state-action $(s(t), a(t))$, the agent selects the action $a(t)$ with the $\epsilon$-greedy policy $\psi$, and then performs the action $a(t)$ and interacts with the environment. Next, the agent will receive the reward $r(t)$ and observe the next state $s(t+1)$ of the environment, meanwhile store the state $\langle s(t), a(t), r(t),s(t+1) \rangle$ into the RB. Every $\kappa$ time steps, the agent updates the DNN by Eq.~\ref{eq:loss} with a mini-batch experience from the replay buffer by means of stochastic gradient descent (SGD). The target net will copy the parameters of the main net by every $\tau$ time steps. During the learning process, we set the learning rate $\sigma$ is 0.001, the $\epsilon$ in $\epsilon$-greedy method is 0.9, the discount accumulative factor $\gamma$ is 0.9, and the step parameters $\tau$ and $\kappa$ are both 2000.

\section{Performance Evaluation}\label{sec:performance}
We evaluate the performance of the proposed DRL-based battery discharging/charging controlling policy through extensive numerical analysis\nop{ and our code is written in Python}.

\subsection{Experiment Setup}

\subsubsection{BS and Power Consumption Data}
In order to show the performance of the proposed method, we mainly consider the 5G BS deployed at the three areas, i.e., \textit{resident area}, \textit{office area}, and \textit{comprehensive area}, whose power consumption within one-week period are illustrated in Fig.~\ref{fig:demand-pattern}, and we assume the power consumption of the same type BSs in different cities (e.g., Beijing, Shanghai and Guangzhou) is the same. For simplicity, we denote the BS deployed at the areas of resident, office, and comprehensive as type I, type II, and type III, respectively. We will apply the BESS aided renewable energy supply solution to different types of BSs in different cities under different weather conditions and evaluate its performance through massive simulation experiment.

\subsubsection{Renewable Energy Generation Data}
\begin{figure*}[!t]
 	\centering
 	\subfigure[The solar PV output power patterns under different weather conditions.]{
 		\includegraphics[width=0.48\textwidth]{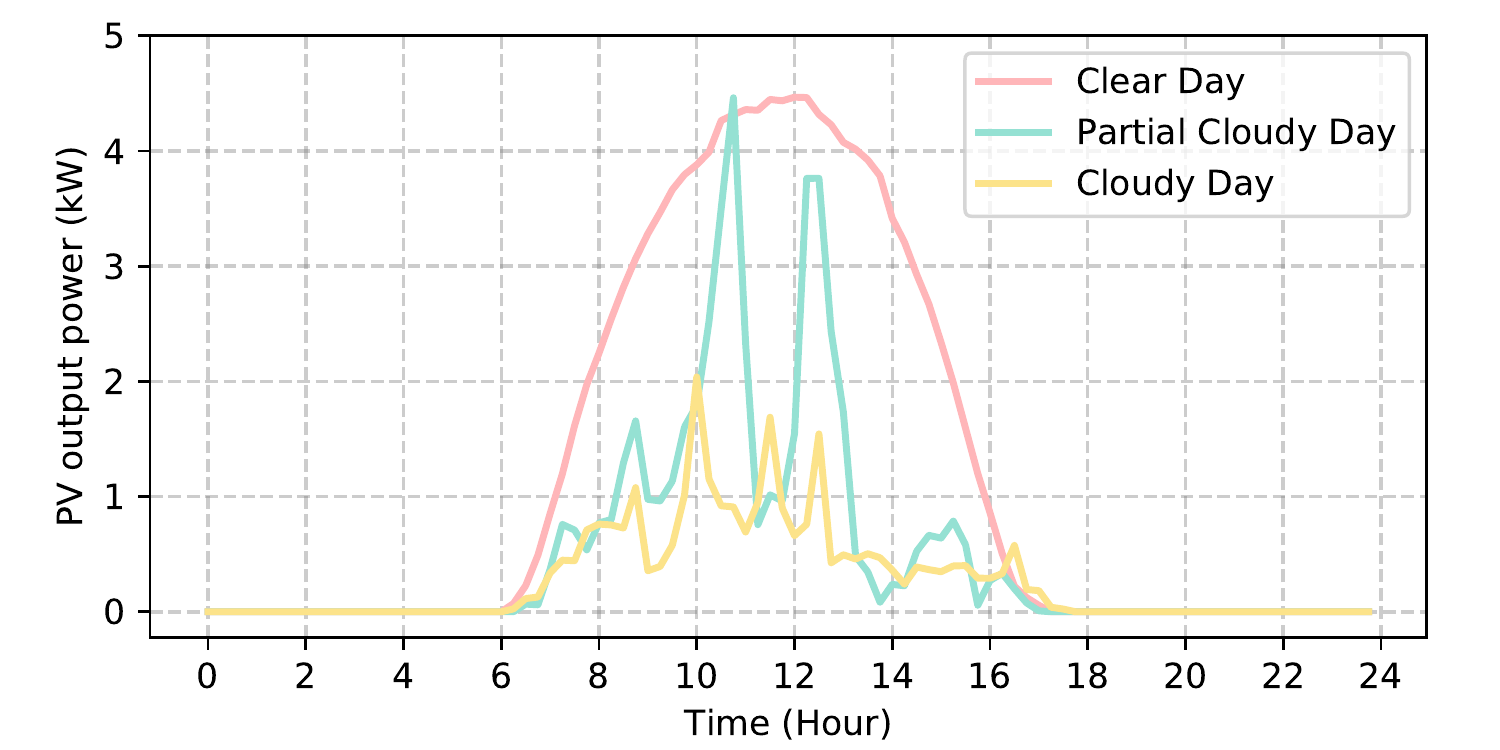}}
 	\hfill
 	\subfigure[The wind turbine output power patterns under different weather conditions.]{
 		\includegraphics[width=0.48\textwidth]{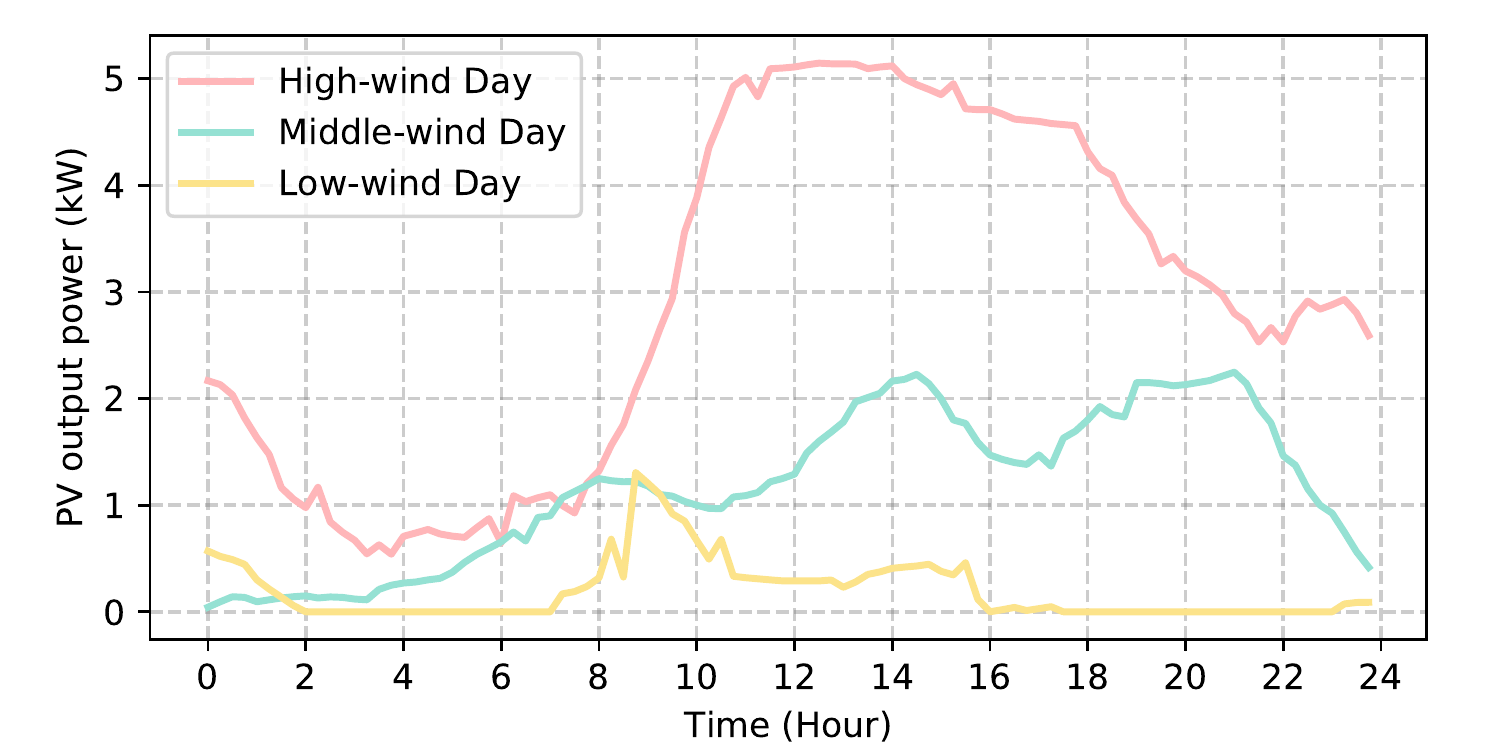}}
 	\caption{(a) The solar PV output power patterns under different weather conditions (i.e., $GHI(t)$, $Temp(t)$, and $ToD(t)$) in one day period. (b) The wind turbine output power patterns under different weather conditions (i.e., $WV(t)$, $WS(t)$, and $HH(t)$) in one day period. }
 	\label{fig:solar-wind-output}
\end{figure*}

In Sec.~\ref{sec:renewable}, we introduce the factors that impact the generation of renewable energy. For simplicity, we divide the weather conditions into three types. Accordingly, the output power pattern of the solar PV and wind turbine could be divided into three types. Specifically, for the solar PV, the weather conditions are divided into the \textit{clear day}, \textit{partial cloudy day}, and \textit{cloudy day}; for the wind turbine, the weather conditions are divided into the \textit{high wind velocity}, \textit{middle wind velocity}, and \textit{low wind velocity}. The output power patterns of the solar PV and wind turbine under different weather conditions are illustrated in Fig.~\ref{fig:solar-wind-output}.

\subsubsection{Equipment Parameter Settings}
In this study, we use a quantity of 15 Panasonic Sc330 solar modules each with a power rating of 330W and JFNH-5kW wind turbine of Qingdao Jinfan Energy Science and Technology Co., Ltd. For the battery storage, we consider the mainstream lithium-ion (LI) battery on the current market. We then refer to~\cite{dabbagh2017shaving, energy-price-contract, us-batter-cost} for parameter settings of electricity billing policy and battery configurations and the main parameter settings are summarized in Table~\ref{tbl:parameters}.

\subsubsection{Scenario Settings}
As the generation of the renewable energy is significantly affected by the weather conditions, we choose three representative cities in China for this paper, i.e., \textit{Beijing}, \textit{Shanghai}, and \textit{Guangzhou}, which has different weather pattern during the billing cycle window (i.e., from 1st June 2020 to 30th June 2020). We compare and analyze the overall energy cost (including energy charge, demand charge and investment cost), detailed controlling results and return of investment (ROI) for three types of BSs (i.e., type I, type II, and type III BSs) in these cities, and the specific day of the weather conditions in these cities during the billing cycle window are shown in Fig.~\ref{fig:weather}. Specifically, i) for \textit{Beijing}, it has more clear days during the billing cycle window, ii) for \textit{Shanghai}, it is in the plum rain season during the billing cycle window, thus it has more high-wind days but less clear days, and iii) for \textit{Guangzhou}, the cloudy days and the low-wind days are relatively more than other two cities.

\renewcommand{\arraystretch}{1.2}
\begin{table}[!t]
    \centering
	\caption{Parameter Settings}\label{tbl:parameters}
	\begin{tabular}{c l l}
		\Xhline{2\arrayrulewidth}
		 		& \emph{Parameter}  & \emph{Setting}  \\ 
		\hline

		\multirow{5}{*}{\shortstack[c]{Billing\\Policy}}
				& billing cycle window $\mathcal{W}$ & one month (30 days) \\
    			& $^{1}$energy charge price $\lambda_{e}$ & US$\$0.049 /kWh$ \\
    			& $^{1}$demand charge price $\lambda_{d}$ & US$\$16.08 /kW$ \\
    			& $^{2}$battery cost $\lambda_{b}$ &  US$\$271/$kWh\\
    	\hline
    	\multirow{4}{*}{\shortstack[c]{Battery\\Config.}}
    			& discharge efficiency $\alpha$    & $85\%$ \\
    			& charge efficiency $\beta$		   &$99.9\%$ \\
    			& max charge rate $R+$ 			& $16$ MW \\
    			& max discharge rate $R-$  		& $8$ MW \\
    	\hline	
    	\multirow{3}{*}{\shortstack[c]{Solar \\ PV}}
    			& power rating $g^s$    & 4950 W\\
    			& price $\lambda_s$		   & US\$3950\\
    			& lifetime $L^s$ 			& 25 years \\
        \hline	
    	\multirow{3}{*}{\shortstack[c]{Wind \\ Turbine}}
    			& power rating $g^w$    & 6000 W\\
    			& price $\lambda_w$		   & US\$4500\\
    			& lifetime $L^w$ 			& 20 years \\
    	
    	\Xhline{2\arrayrulewidth}  	
	\end{tabular}

	\begin{flushleft}
	$^{1}$Prices of energy/demand charges in 2018, referring to the contract in~\cite{energy-price-contract}. 
	
	$^{2}$Battery capacity costs in 2018, referring to the data in~\cite{us-batter-cost}.
	\end{flushleft}
	\vspace{-0.15in}
\end{table}

\subsection{Performance under Different Weather Conditions}
As is shown in Fig.~\ref{fig:solar-wind-output}, the output power patterns of the solar PV and wind turbine are both divided into three types under different weather conditions. Accordingly, the weather pattern can be divided into nine types: \textit{clear \& high-wind day}, \textit{clear \& middle-wind day}, \textit{clear \& low-wind day}, \textit{partial cloudy \& high-wind day}, \textit{partial cloudy \& middle-wind day}, \textit{partial cloudy \& low-wind day}, \textit{cloudy \& high-wind day}, \textit{cloudy \& middle-wind day}, and \textit{cloudy \& low-wind day}.

\begin{figure*}[!t]
 	\centering
 	 	\subfigure[The power supply pattern under the clear \& high-wind day.]{
 		\includegraphics[width=0.32\textwidth]{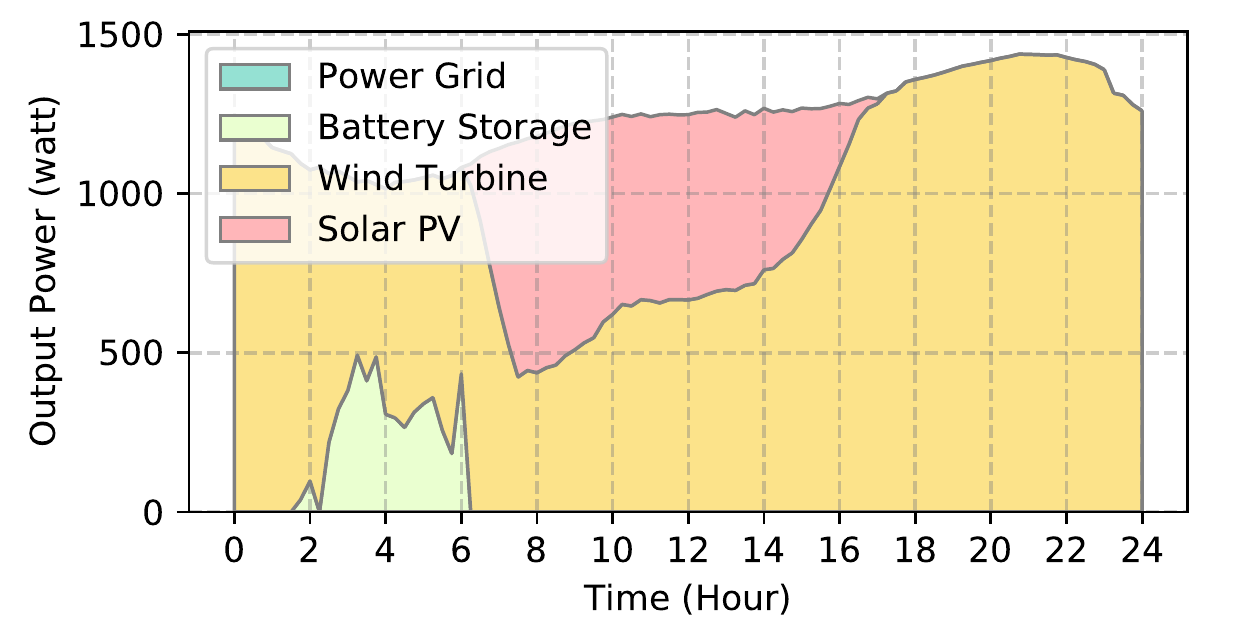}}
 	\hfill
 	\subfigure[The power supply pattern under the clear \& middle-wind day]{
 		\includegraphics[width=0.32\textwidth]{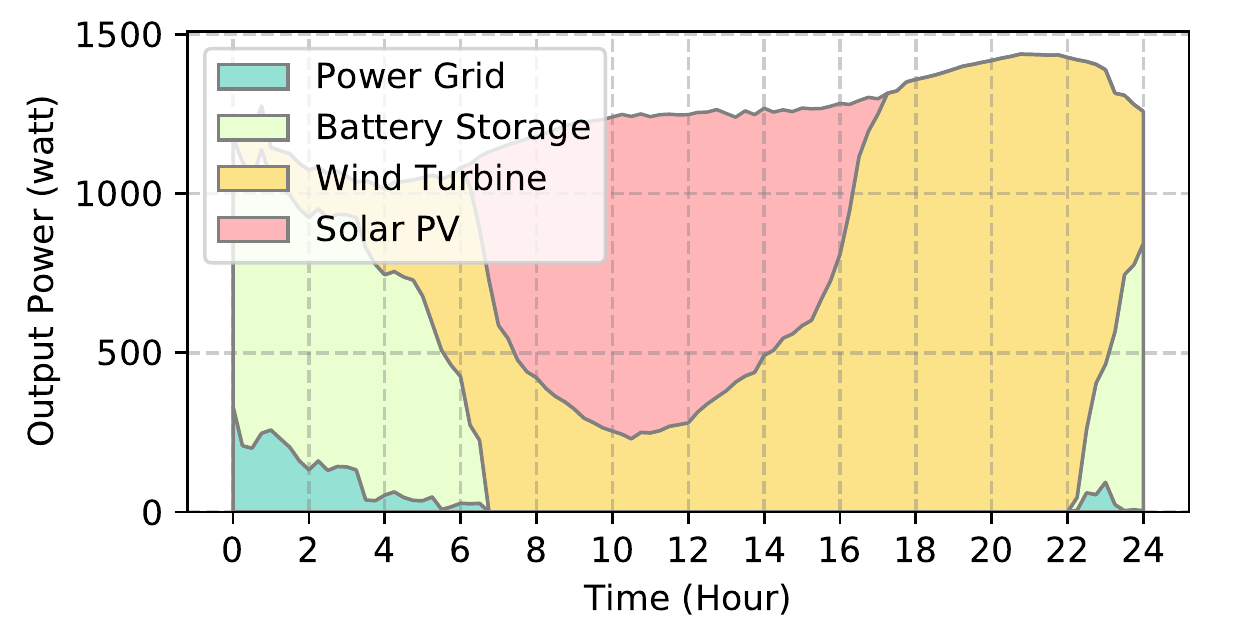}}
 	\hfill
 	\subfigure[The power supply pattern under the clear \& low-wind day]{
 		\includegraphics[width=0.32\textwidth]{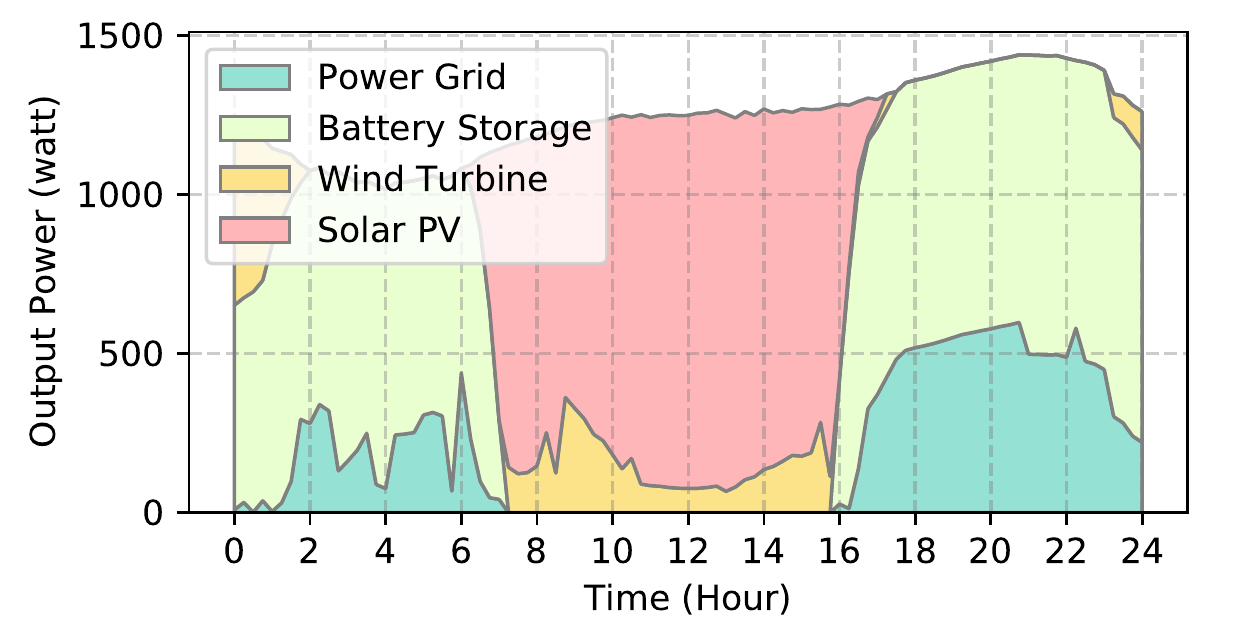}}
 	\\
 	 	\subfigure[The power supply pattern under the partial cloudy \& high-wind day]{
 		\includegraphics[width=0.32\textwidth]{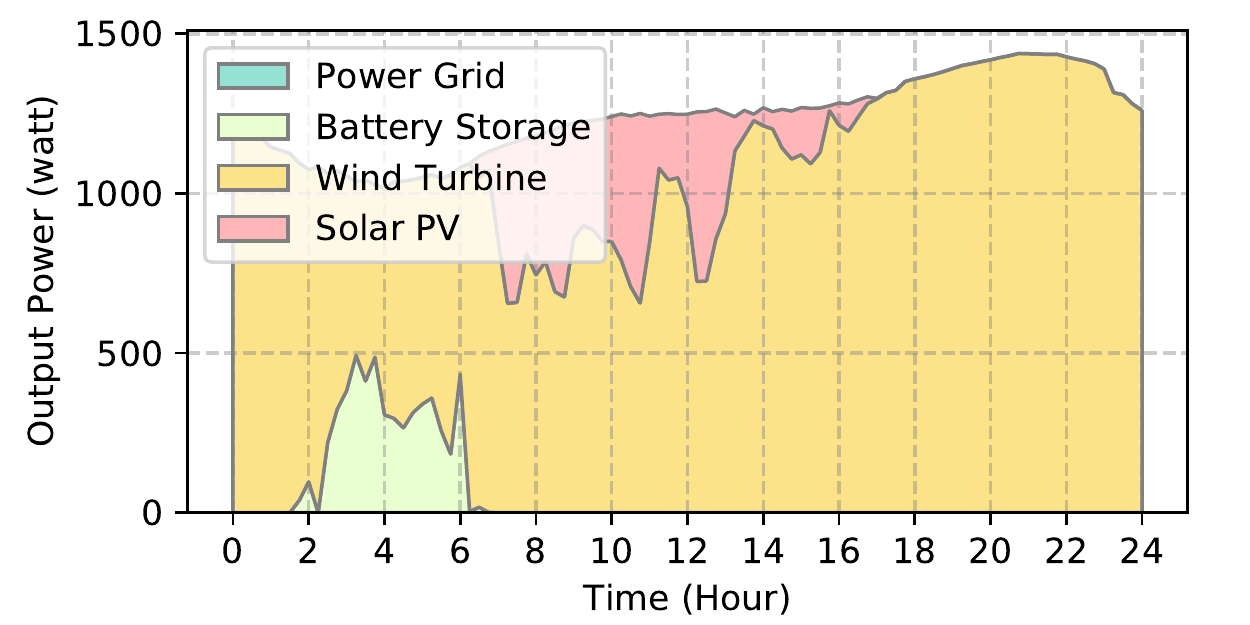}}
 	\hfill
 	\subfigure[The power supply pattern under the partial cloudy \& middle-wind day]{
 		\includegraphics[width=0.32\textwidth]{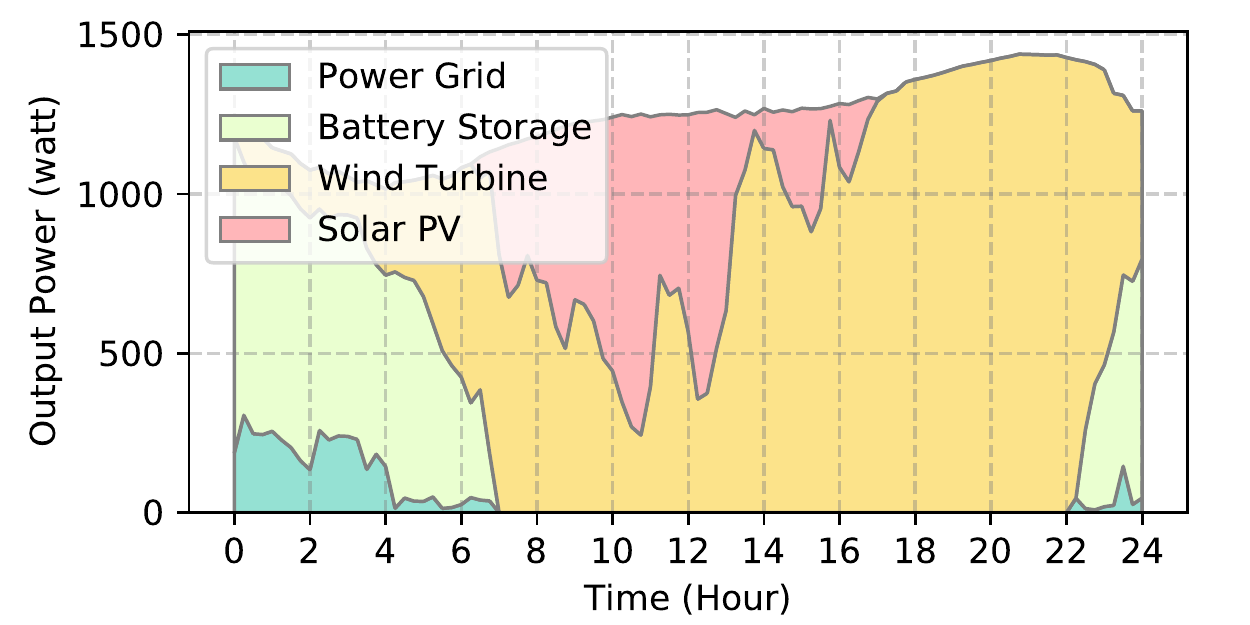}}
 	\hfill
 	\subfigure[The power supply pattern under the partial cloudy \& low-wind day]{
 		\includegraphics[width=0.32\textwidth]{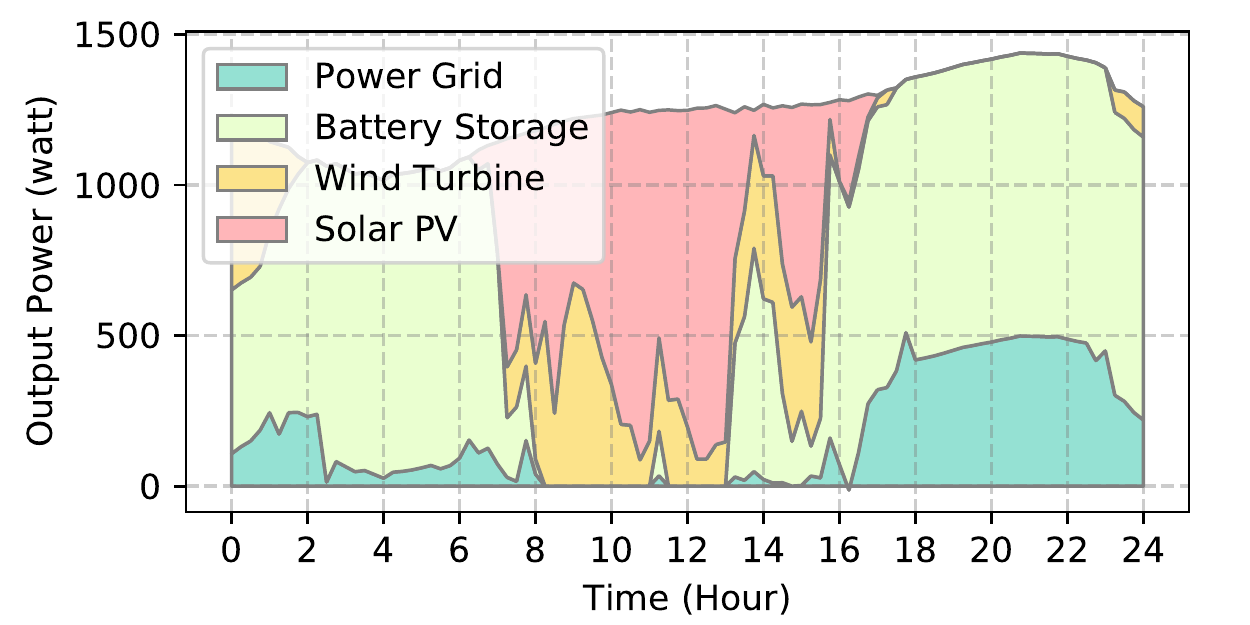}}
 	\\
 	\subfigure[The power supply pattern under the cloudy \& high-wind day]{
 		\includegraphics[width=0.32\textwidth]{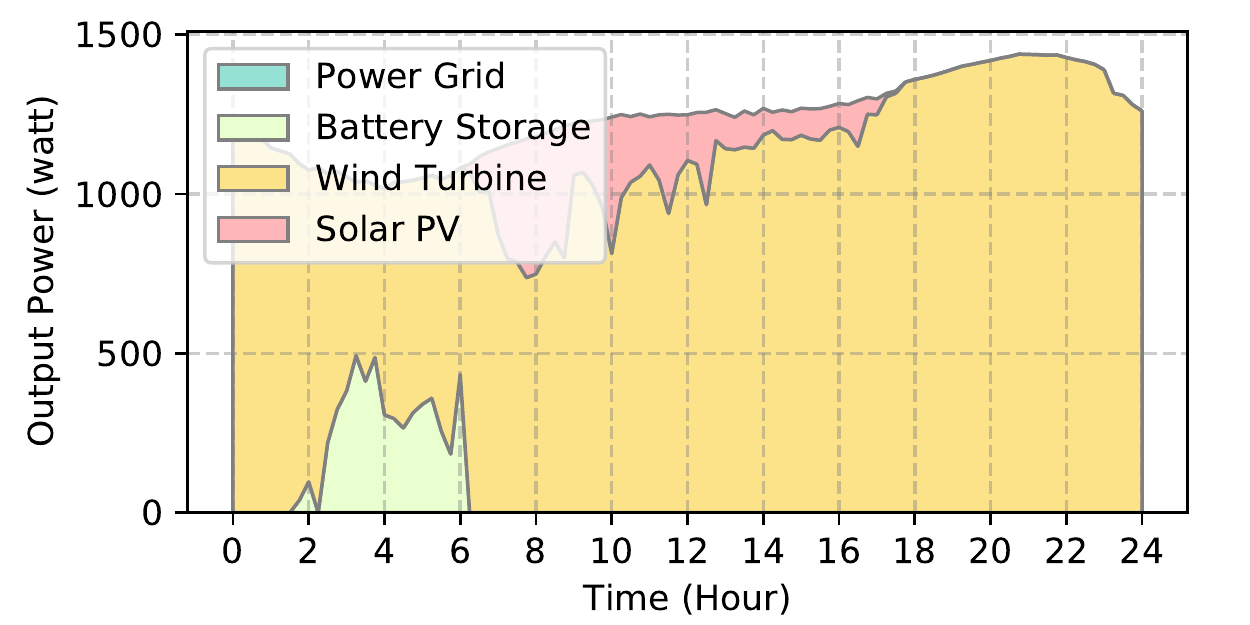}}
 	\hfill
 	\subfigure[The power supply pattern under the cloudy \& middle-wind day]{
 		\includegraphics[width=0.32\textwidth]{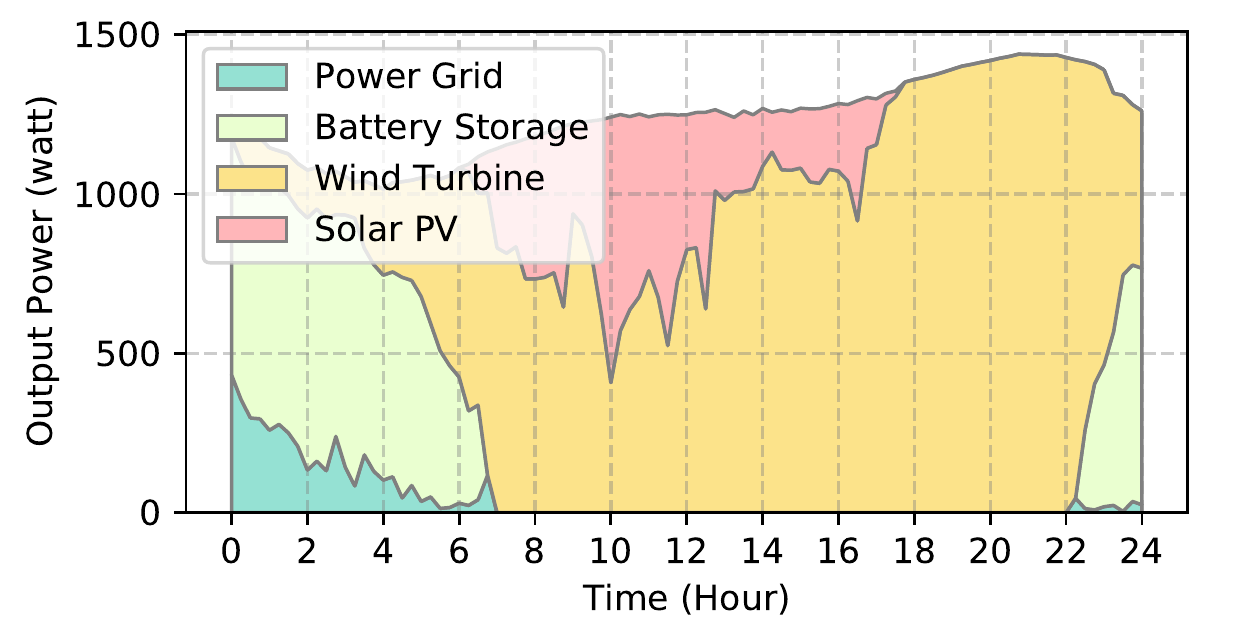}}
 	\hfill
 	\subfigure[The power supply pattern under the cloudy \& low-wind day]{
 		\includegraphics[width=0.32\textwidth]{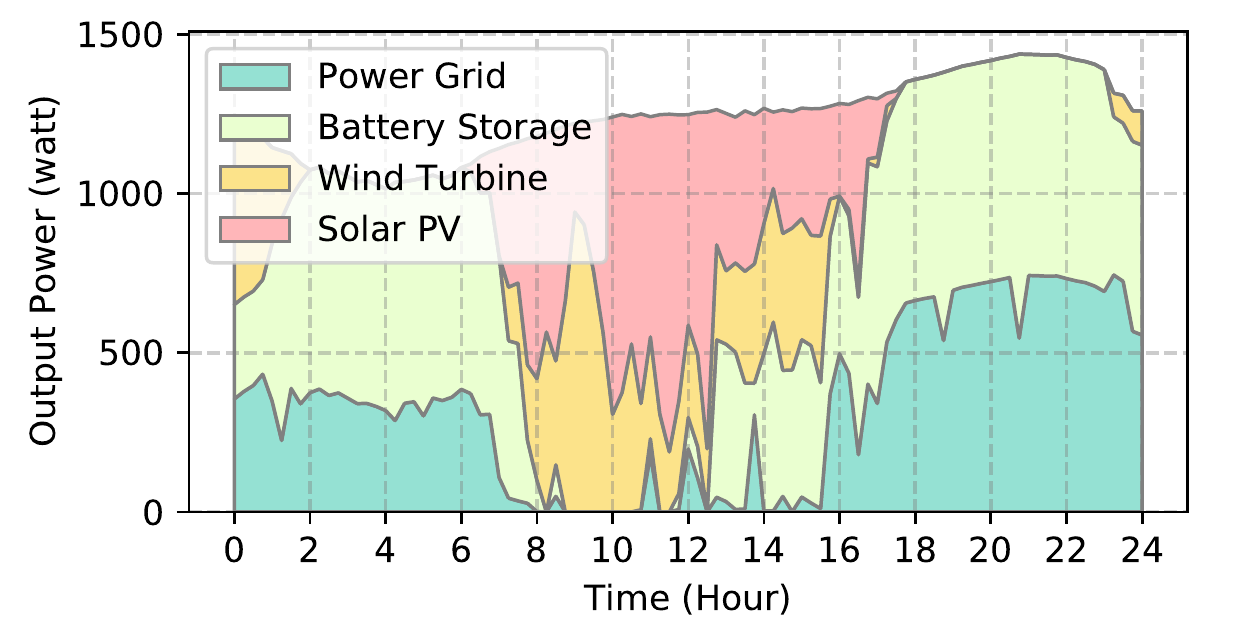}}
 	\caption{The power supply pattern of a single 5G BS at area of resident is supplied by different power supply methods under different weather conditions in one day period. }
 	\label{fig:power-pattern-resident}
 	\vspace{-0.15in}
\end{figure*}

The power supply patterns under different weather conditions in one day period of 5G BS at the area of resident, office, and comprehensive are illustrated in Fig.~\ref{fig:power-pattern-resident}, Fig.~\ref{fig:power-pattern-office}, and Fig.~\ref{fig:power-pattern-comprehensive} (in the appendix), respectively. As we can see, the BESS aided renewable energy supply solution could significantly reduce the power from the grid (i.e., \textit{energy charge} and \textit{demand charge}). Specifically, with the increase of radiation and wind velocity, renewable energy generation increased accordingly. It could cover most of the power demand and reduce the power supplied from the power grid. Especially, under high-wind days, the power demand could be totally supplied by the renewable energy and battery storage and need 0 power from the grid. 

\begin{figure}[t!]
    \centering
	\centerline{\includegraphics[width=1\linewidth]{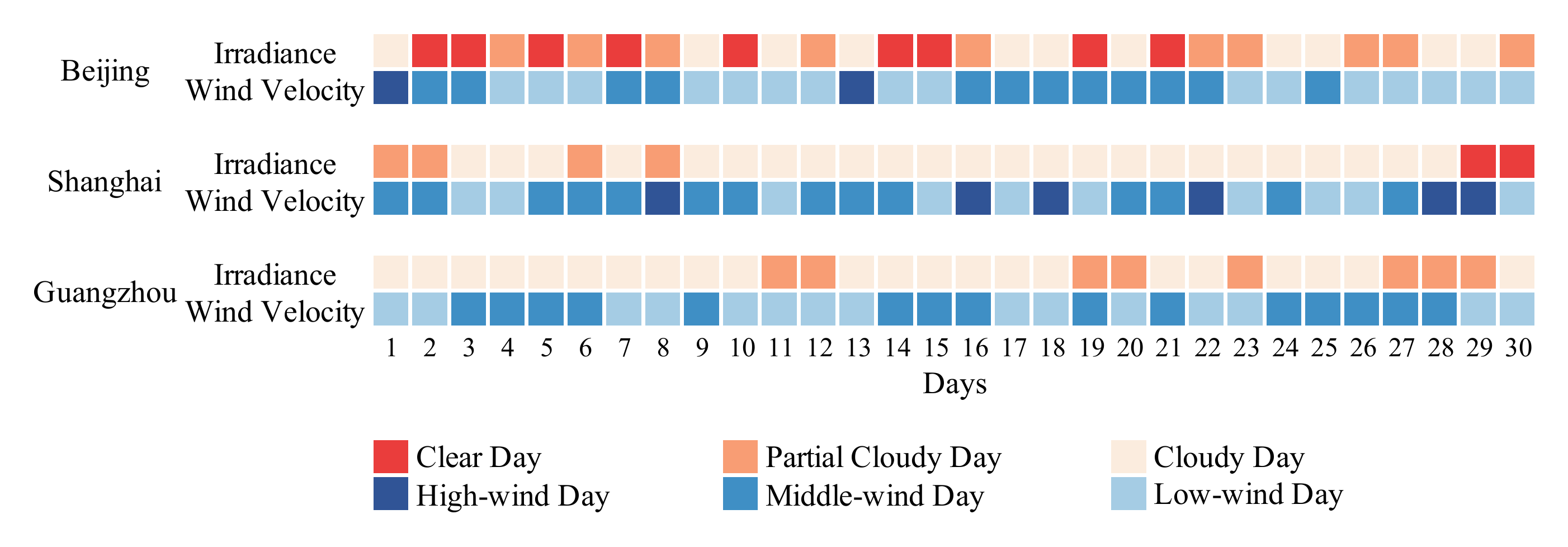}}
	\caption{The weather data is obtain from~\cite{china-weather}, and the billing cycle window is from 1st June 2020 to 30th June 2020.}\label{fig:weather}
	\setlength{\abovecaptionskip}{0cm}
	\vspace{-0.15in}
\end{figure}

After calculating the power supply paradigm under different weather patterns, we can derive the electricity bill of these three types of BSs during the billing cycle in different cities (i.e., different weather patterns, which is illustrated in Fig.~\ref{fig:weather}), and the results from all the set scenarios are summarized in Table~\ref{tbl:results}.

\renewcommand{\arraystretch}{1.2}
\begin{table*}[!t]
\centering
\caption{Results Summary (One Billing Cycle)}\label{tbl:results}
\begin{adjustbox}{width=1\linewidth}
\begin{tabular}{c c c c c c c c}
\Xhline{2\arrayrulewidth}
\textbf{BS Type} & \textbf{Scenario} & \multicolumn{1}{c}{\textbf{Energy Charge (\$)}} & \multicolumn{1}{c}{\textbf{Demand Charge (\$)}} & \multicolumn{1}{c}{\textbf{Investment Cost (\$)}} & \multicolumn{1}{c}{\textbf{Cost Saving (\$)}} & \multicolumn{1}{c}{\textbf{Saving Ratio (\%)}}  \\
\hline

\multirow{4}{*}{Type I}
&No deployment & 44.6 & 23.1 & 0 & / & /  \\

&Deployment in Beijing &5.0 &12.0 &0.4 &50.4 &74.4 \\

&Deployment in Shanghai &4.7 &12.0 &0.4 &50.7 &74.8  \\

&Deployment in Guangzhou &5.9 &12.0 &0.3 &49.5 &73.2  \\

\hline

\multirow{4}{*}{Type II}
&No deployment & 40.1 & 20.2 & 0 & / & /  \\

&Deployment in Beijing &4.8 &9.1 &0.3 &46.1 &76.4 \\

&Deployment in Shanghai &3.8 &9.1 &0.4 &47.0 &77.9  \\

&Deployment in Guangzhou &5.3 &9.1 &0.3 &45.6 &75.6  \\

\hline

\multirow{4}{*}{Type III}
&No deployment & 45.6 & 22.8 & 0 & / & /  \\

&Deployment in Beijing &6.8 &13.9 &0.3 &47.4 &69.3 \\

&Deployment in Shanghai &5.7 &13.9 &0.4 &48.4 &70.8  \\

&Deployment in Guangzhou &7.9 &13.9 &0.2 &46.4 &67.8  \\

\Xhline{2\arrayrulewidth}
\end{tabular}
\end{adjustbox}
\vspace{-0.15in}
\end{table*}

Specifically, for a single 5G BS without the proposed power supply paradigm, the energy charge and the demand charge are {\$45.6} and {\$22.8}, respectively. However, after utilizing the BESS aided renewable energy supply solution on the 5G BS, the electricity bill is significantly reduced. Especially in {Shanghai}, which has relatively more clear and high-wind days, the energy charge and the demand charge can be reduced to {\$3.8} and {\$9.1}, respectively. Although there exists equipment degradation during the discharge/charge cycles, the investment cost still keeps at a well accepted level. The highest cost saving for the BS which utilized the proposed power supply paradigm in Beijing, Shanghai, and Guangzhou in one billing cycle is {\$50.4}, {\$50.7} and {\$49.5}, respectively. Accordingly, the saving ratio can be up to {74.4\%}, {74.8\%} and {73.2\%}, respectively.

\subsection{Performance under Different Types of BSs}
As the different types of BSs has diverse power demand, resulting in different energy charge and demand charge, thus the performance of deployment of the BESS aided renewable energy supply solution could be different.

Specifically, as is shown in Table~\ref{tbl:results}, the type I BS has the highest cost saving compared to other two types of BSs, i.e., {\$50.4} in Beijing, {\$50.7}, and {\$49.5}. This is because that type I BS has the biggest power demand and peak value (near 1450 watts), making it has great potential in energy-saving and peak power shaving. Besides, as type II BS's power demands are relatively small, the generated and stored renewable energy can effectively reduce the power grid supply. Therefore it has the highest saving ratio, i.e., {76.4\%} in Beijing, {77.9\%} in Shanghai, and {75.6\%} in Guangzhou.

\subsection{ROIs of Different City and Type Deployment}
The return of investment (ROI) is a financial metric defined by the benefit (cost saving in our case) divided by the total investment. It indicates the probability of gaining a return from an investment and has been widely used to evaluate the efficiency of an investment~\cite{roi-wiki}. Typically, a bigger ROI value indicates a higher investment efficiency. With the costs of renewable energy generator and battery storage (given in Table~\ref{tbl:parameters}), the total investments can be calculated. Accordingly, the ROIs can thus be derived with the results in Table~\ref{tbl:results}.

\renewcommand{\arraystretch}{1.2}
\begin{table}[h]
    \centering
	\caption{Parameter Settings}\label{tbl:roi}
	\begin{tabular}{c c c c}
		\Xhline{2\arrayrulewidth}
		\textbf{BS Type} 		& \textbf{Beijing} & \textbf{Shanghai} & \textbf{Guangzhou} \\ 
		\hline
    	Type I &5.43\% &5.46\%  &5.33\%  \\
    	Type II &4.97\%  &5.06\%  &4.91\%  \\
    	Type III &5.11\%  &5.21\%  &5.00\%  \\
    	\Xhline{2\arrayrulewidth}  	
	\end{tabular}
\end{table}

The ROIs of different types of BSs deployed in different cities are shown in Table~\ref{tbl:roi}. Specifically, type I BS has the highest ROI, which could reach to {5.43\%} in Beijing, {5.46\%} in Shanghai, and {5.33\%} in Guangzhou, respectively, indicating a relatively high investment efficiency for the operators. This is because that type I BS has the biggest cost saving. 

As the equipment's cost is estimated to decrease dramatically in the future~\cite{estimate-LI-cost}, and the ROI could rise significantly in 5G and beyond. Additionally, as we can see, the city with more clear and high-wind days will obtain a bigger ROI value, thus the proposed solution is more suitable for those cities with more sunny and windy days.

It is worth noting that, we assume the deployed renewable energy generator and the battery storage only supply power to one single 5G BS, and thus the surplus renewable energy (when the battery is full) will be discarded. This actually leads to a relatively low utilization, as given in this work. In practice, the generated renewable energy could supply to multiple BSs~\cite{tang2020shiftguard}, so that the ROI and utilization of the renewable energy could be further improved.

\section{Related Work}\label{sec:relatedwork}
The most involved related literatures can be divided into the following three categories. \nop{including the base station energy-saving method, general system peak power shaving and battery storage optimal control.}

\subsection{Base Station Energy-saving Method}
With the increase of the BS power consumption, the energy-efficient design of cellular networks has recently received significant attention. Typically, the BS energy-saving methods are divided into three levels, \textit{equipment-level energy saving}, \textit{site-level energy saving} and \textit{network-level energy saving}~\cite{china-mobile}.

\textbf{On the equipment-level}, researchers propose new schemes (e.g., the scheme in this paper), new materials (e.g., semiconductor material), and new functions so as to achieve the energy-efficient goal. Besides, some liquid heat dissipation, high power amplifier efficiency, and high integration applications are applied in the BS so as to reduce the power consumption of the whole machine year by year.

\textbf{On the site-level}, one common scheme is to switching-on/off the BS related with the traffic load~\cite{chiaraviglio2008energy, oh2010energy, kumar2015energy}. To be detail, switch-on the BS when the traffic load is large, and switch-off the BS when the traffic load is low. In addition, by combining with AI, the accuracy of the traffic load prediction can be improved so that the corresponding energy-saving policies can be elaborately formulated.

\textbf{On the network-level}, With the deployment of the 5G network, multiple networks (e.g., 4G/LTE and 5G) coexist in the current network, so the energy efficiency can be improved through the application of network-level energy-saving technology. Based on the basic data (e.g., configuration and performance) of the network and the built-in strategy, the multi network cooperative energy-saving technology~\cite{chen2011network} can realize the goal of reducing energy consumption by turning off the BS under the condition of ensuring service quality.

\subsection{General System Peak Power Shaving}
Peak power (i.e., the demand charge) is a sensitive factor for the power grid, as it occurs occasionally and takes place only for a small percentage of the time in a day~\cite{uddin2018review}. The traditional solution is to increase the grid capacity, leading to uneconomic and inefficiency to supply peak power. The peak power shaving is a preferable approach to overcome these disadvantages, making the load curve flatten by reducing the peak amount of load and shifting it to times of lower load~\cite{nourai2008load}. Typically, the related works are divided into the following three categories:

\textbf{Peak shaving using energy storage system (ESS)}: Integrating energy storage systems to the grid is the most potent strategy of peak shaving due to its economic benefits~\cite{reihani2016load, son2014real, lavrova2012analysis}. Specifically, peak power shaving is achieved through the process of charging ESS when demand is low (off-peak period) and discharging when demand is high (one-peak period).

\textbf{Peak shaving using electric vehicles (EV)}: Since the storage energy of electric vehicles is usually not fully utilized each day, this technology has the potential to provide peak shaving service~\cite{tse2016use, yao2014optimization, leemput2012case}. For example, Alam et al.~\cite{alam2014controllable} proposed an effective strategy to utilize PEV batteries for both traveling and peak shaving purpose.

\textbf{Peak shaving using demand side management (DSM)}: Demand side management refers to the programs that may influence the customers to balance their electricity consumption with the power supply system's generation capacity~\cite{crosbie2017future,mohagheghi2010demand,muratori2015residential,samanta2017energy}. For example, Rozali et al.~\cite{rozali2015peak} aimed to achieve maximum peak shaving through DR under power shaving analysis.

\subsection{Battery Storage Optimal Control}
The optimal control of energy storage has been extensively studied in the past. Most related works formulate an optimization problem that aims to maximize the revenue generated by the battery storage co-located with renewable energy generator.

Babacan et al.~\cite{babacan2017distributed} proposed a convex program to minimize the electricity bill of operators. Ratnam et al.~\cite{ratnam2015optimization} aimed to maximize the daily operational savings that accrue to customers while penalizing large voltage swings stemming from reverse power flow and peak load. Kazhamiaka et al.~\cite{kazhamiaka2017influence} studied the profitability of residential PV-storage systems in three jurisdictions and set up an integer linear program to determine the battery controlling policy. These works assume the generations of renewable energy and the power demand are known in advance and can be optimized in an offline way. However, these assumptions are unpractical in the real world. 

Several papers study the optimal control of batteries under uncertainty and randomness. Guan et al.~\cite{guan2015reinforcement} utilized a reinforcement learning method to minimize the homeowner’s cost by taking an action that yields the best expected reward. EnergyBoost~\cite{qi2019energyboost} could provide a predictable ability of the renewable energy generation and power demand. However, these works are only applied in the home scenario, which generates a few demands compared to 5G BS. Therefore, we propose the DRL-based method to tackle the problem of large and constrained state- and action-space and the uncertainty of renewable energy generation and power demand.
 
\section{Conclusions}\label{sec:conclusion}
To copy with the ever-increasing electricity bill for mobile operators in 5G era, we proposed a BESS aided reconfigurable energy supply solution for the 5G BS system, which models the battery discharging/charging reconfiguration as an optimization problem. With our proposed solution, besides the power grid, a BS can be powered by the renewable energy and the battery storage, to cut down the total energy cost. To solve the problem under the dynamic power demands and renewable energy generation, we developed a DRL-based approach to the BESS operation that accommodates all factors in the modeling phase and makes decisions in real-time. To evaluate the performance of our solution, we chose three cities with different weather patterns for experiments. The experimental results show that our reconfigurable power supply solution can significantly reduce the electricity bill and improve the renewable energy utilization.

\balance{\bibliographystyle{IEEEtran}
\bibliography{main}}

\vskip -1\baselineskip plus -1fil

\begin{IEEEbiographynophoto}
{Hao~Yuan} received the B.S. degree in management science and engineering from National University of Defense Technology, Changsha, China, in 2019. He is currently working towards the M.S. degree in the same department. His main research interests include edge computing and green communication.
\end{IEEEbiographynophoto}

\vskip -1\baselineskip plus -1fil

\begin{IEEEbiographynophoto}
{Guoming~Tang} is a research fellow at the Peng Cheng Laboratory, Shenzhen, Guangdong, China. He received his Ph.D. degree in Computer Science from the University of Victoria, Canada, in 2017, and the Bachelor's and Master's degrees from the National University of Defense Technology, China, in 2010 and 2012, respectively. He was also a visiting research scholar of the University of Waterloo, Canada, in 2016. His research mainly focuses on green computing, computing for green and edge computing.
\end{IEEEbiographynophoto}

\vskip -1\baselineskip plus -1fil

\begin{IEEEbiographynophoto}
{Deke~Guo} received the B.S. degree in industry engineering from the Beijing University of Aeronautics and Astronautics, Beijing, China, in 2001, and the Ph.D. degree in management science and engineering from the National University of Defense Technology, Changsha, China, in 2008. He is currently a Professor with the College of System Engineering, National University of Defense Technology, and is also with the College of Intelligence and Computing, Tianjin University. His research interests include distributed systems, software-defined networking, data center networking, wireless and mobile systems, and interconnection networks. He is a senior member of the IEEE and a member of the ACM.
\end{IEEEbiographynophoto}

\vskip -1\baselineskip plus -1fil

\begin{IEEEbiographynophoto}
{Kui~Wu} received the BSc and the MSc degrees in computer science from Wuhan University, China, in 1990 and 1993, respectively, and the PhD degree in computing science from the University of Alberta, Canada, in 2002. He joined the Department of Computer Science, University of Victoria, Canada, in 2002, where he is currently a Full Professor. His research interests include smart grid, mobile and wireless networks, and network performance evaluation. He is a senior member of the IEEE.
\end{IEEEbiographynophoto}

\vskip -1\baselineskip plus -1fil

\begin{IEEEbiographynophoto}
{Xun~Shao} received his Ph.D. in information science from the Graduate School of Information Science and Technology, Osaka University, Japan, in 2013. From 2013 to 2017, he was a researcher with the National Institute of Information and Communications Technology (NICT) in Japan. Currently, he is an Assistant Professor at the School of Regional Innovation and Social Design Engineering, Kitami Institute of Technology, Japan. His research interests include distributed systems and networking. He is a member of the IEEE and IEICE.
\end{IEEEbiographynophoto}

\vskip -1\baselineskip plus -1fil

\begin{IEEEbiographynophoto}
{Keping Yu} received the M.E. and Ph.D. degrees from the Graduate School of Global Information and Telecommunication Studies, Waseda University, Tokyo, Japan, in 2012 and 2016, respectively. He was a Research Associate and a Junior Researcher with the Global Information and Telecommunication Institute, Waseda University, from 2015 to 2019 and 2019 to 2020, respectively, where he is currently an Assistant Professor. His research interests include smart grids, information-centric networking, the Internet of Things, artificial intelligence, blockchain, and information security. He is a Member of the IEEE.
\end{IEEEbiographynophoto}

\vskip -1\baselineskip plus -1fil

\begin{IEEEbiographynophoto}
{Wei Wei} received the M.S. and Ph.D. degrees from Xi’an Jiaotong University, Xi’an, China, in 2005 and 2011, respectively. He is currently an Associate Professor with the School of Computer Science and Engineering, Xi’an University of Technology, Xi’an. He ran many funded research projects as a Principal Investigator and Technical Member. He has published over 100 research articles in international conferences and journals. His current research interests include the area of wireless networks, wireless sensor networks application, image processing, mobile computing, distributed computing, and pervasive computing, the Internet of Things, and sensor data clouds. He is a Senior Member of the China Computer Federation. He is an Editorial Board Member of the Future Generation Computer System, the IEEE Access, Ad Hoc \& Wireless Sensor Network, Institute of Electronics, Information and Communication Engineers, and KSII Transactions on Internet and Information Systems.
\end{IEEEbiographynophoto}

\appendices


\section{Results from Office \& Comprehensive Areas}

\begin{figure*}[!h]
 	 	\subfigure[The power supply pattern under the clear \& high-wind day.]{
 		\includegraphics[width=0.32\textwidth]{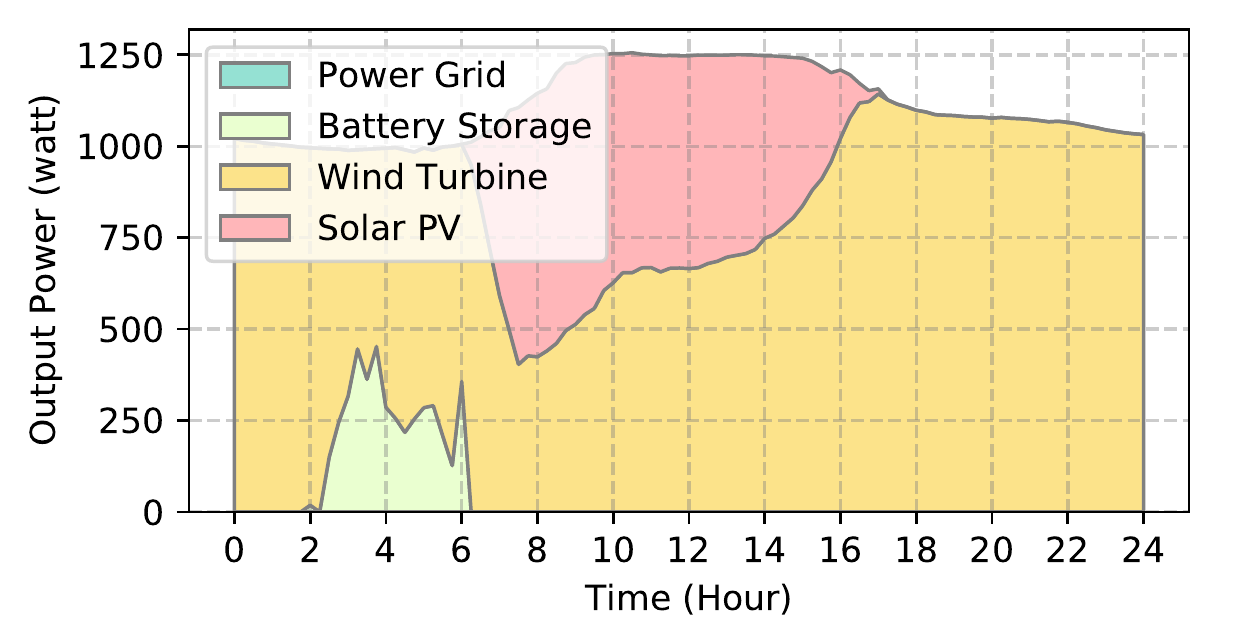}}
 	\hfill
 	\subfigure[The power supply pattern under the clear \& middle-wind day]{
 		\includegraphics[width=0.32\textwidth]{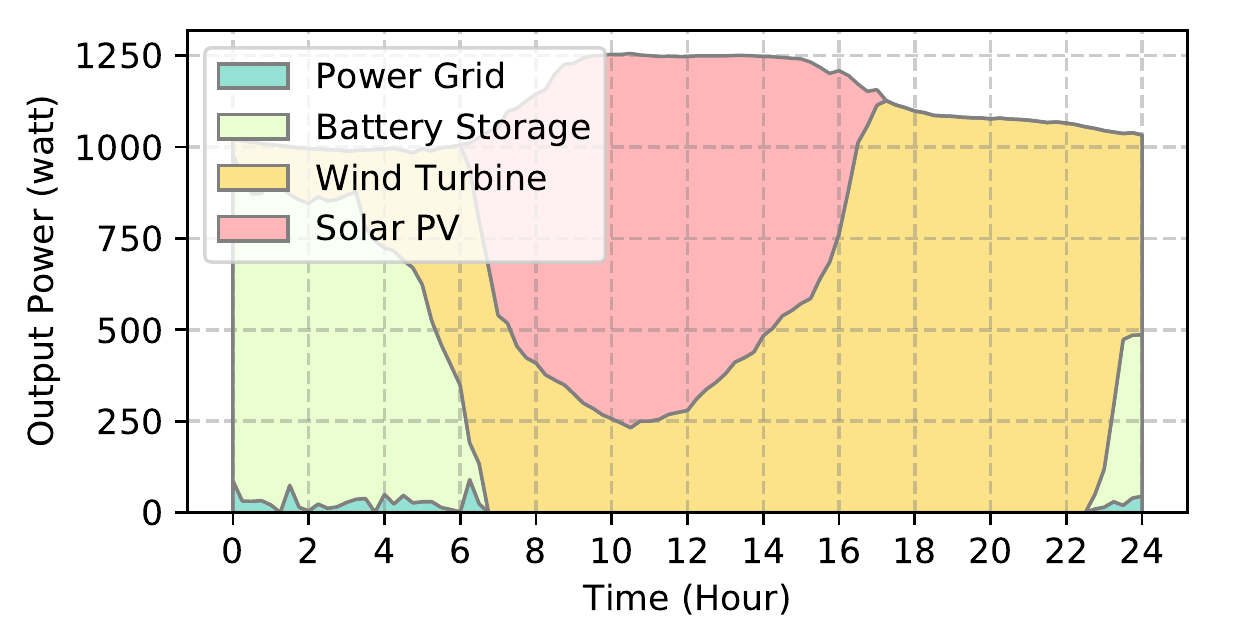}}
 	\hfill
 	\subfigure[The power supply pattern under the clear \& low-wind day]{
 		\includegraphics[width=0.32\textwidth]{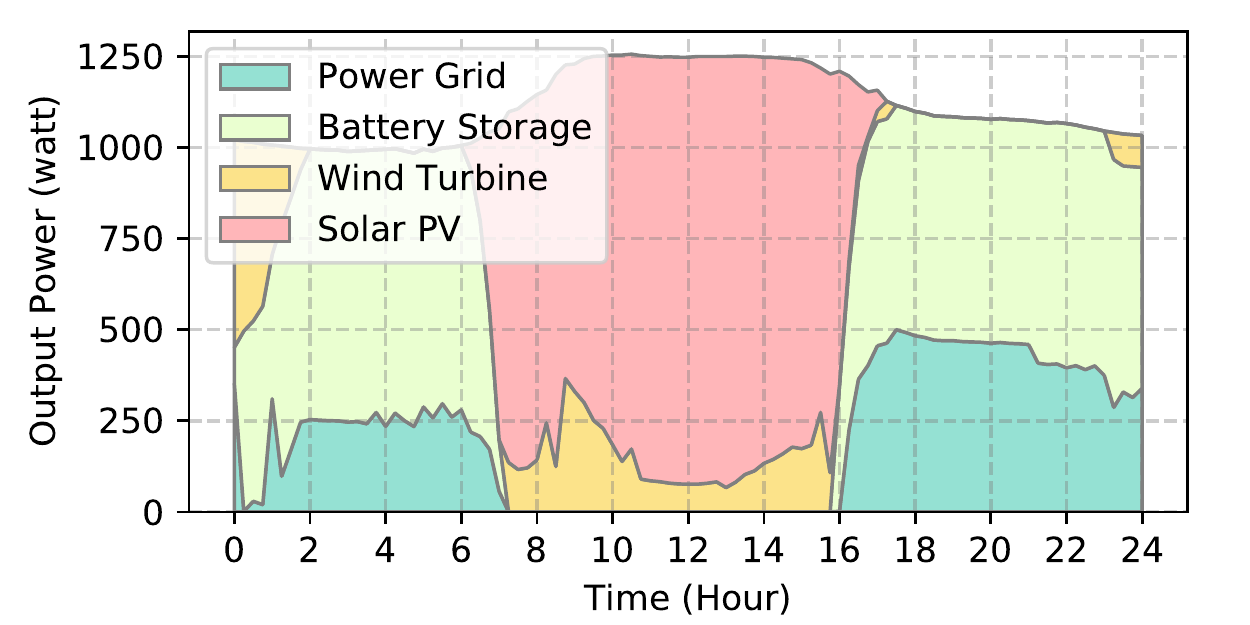}}
 	\\
 	 	\subfigure[The power supply pattern under the partial cloudy \& high-wind day]{
 		\includegraphics[width=0.32\textwidth]{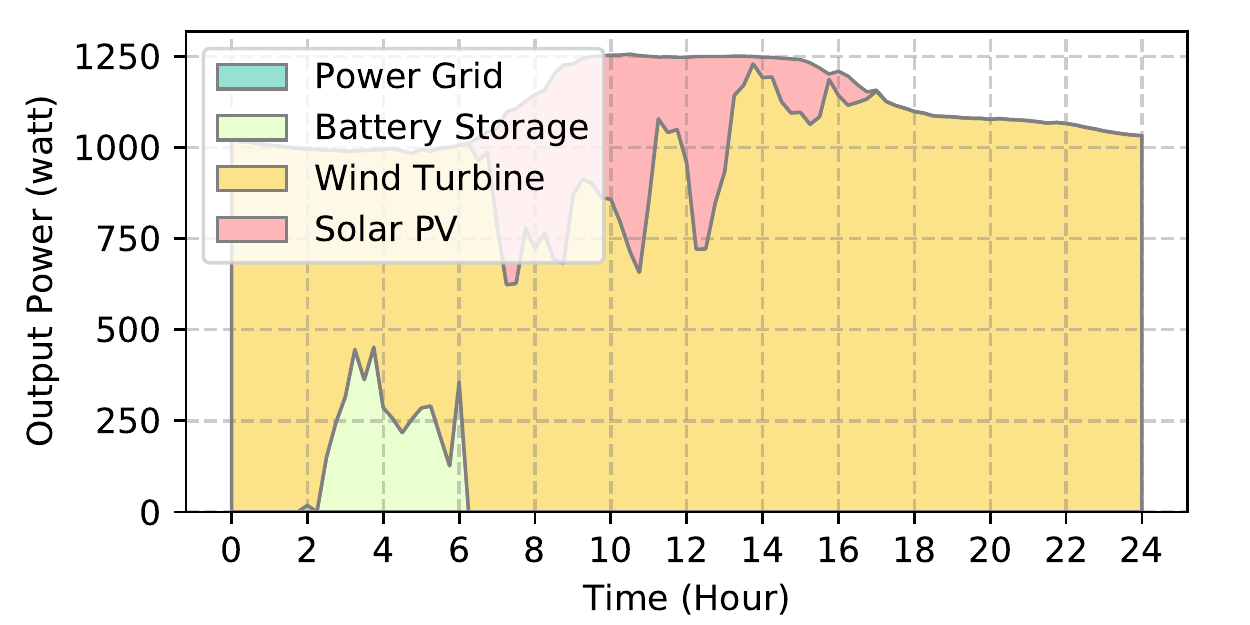}}
 	\hfill
 	\subfigure[The power supply pattern under the partial cloudy \& middle-wind day]{
 		\includegraphics[width=0.32\textwidth]{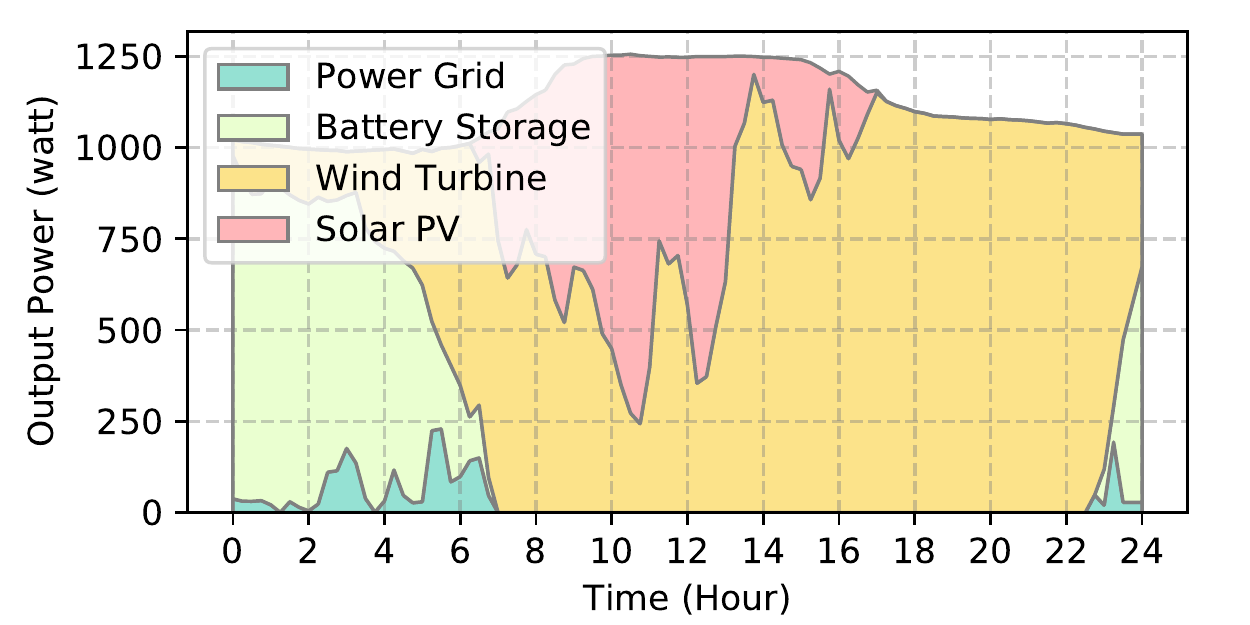}}
 	\hfill
 	\subfigure[The power supply pattern under the partial cloudy \& low-wind day]{
 		\includegraphics[width=0.32\textwidth]{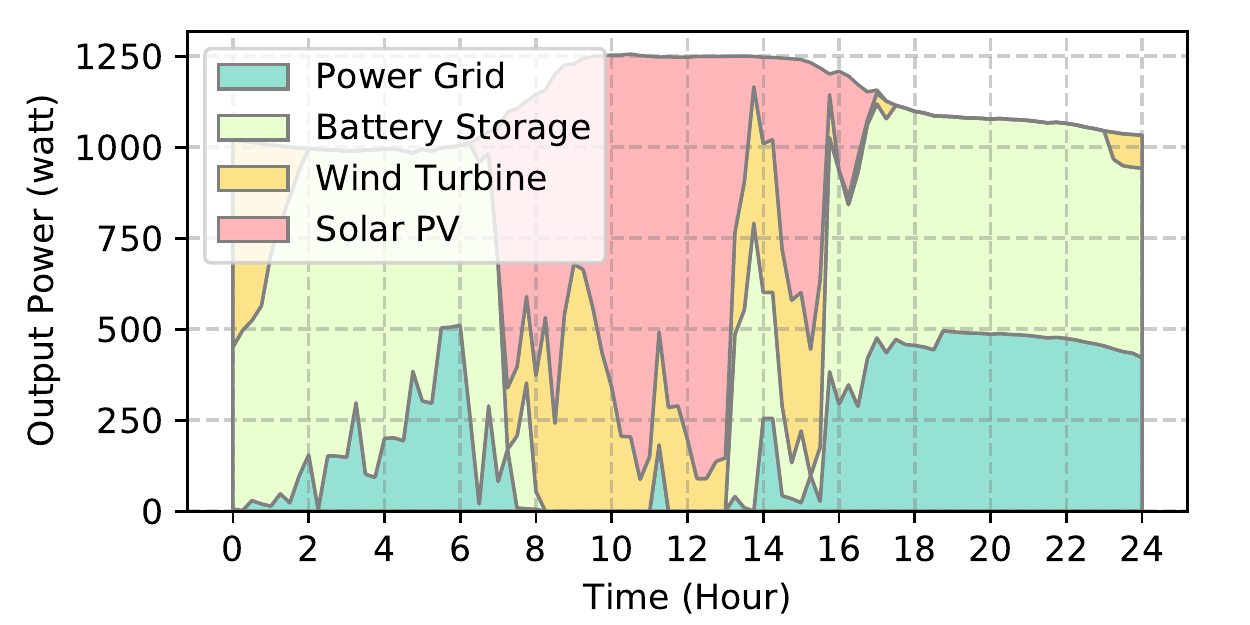}}
 	\\
 	\subfigure[The power supply pattern under the cloudy \& high-wind day]{
 		\includegraphics[width=0.32\textwidth]{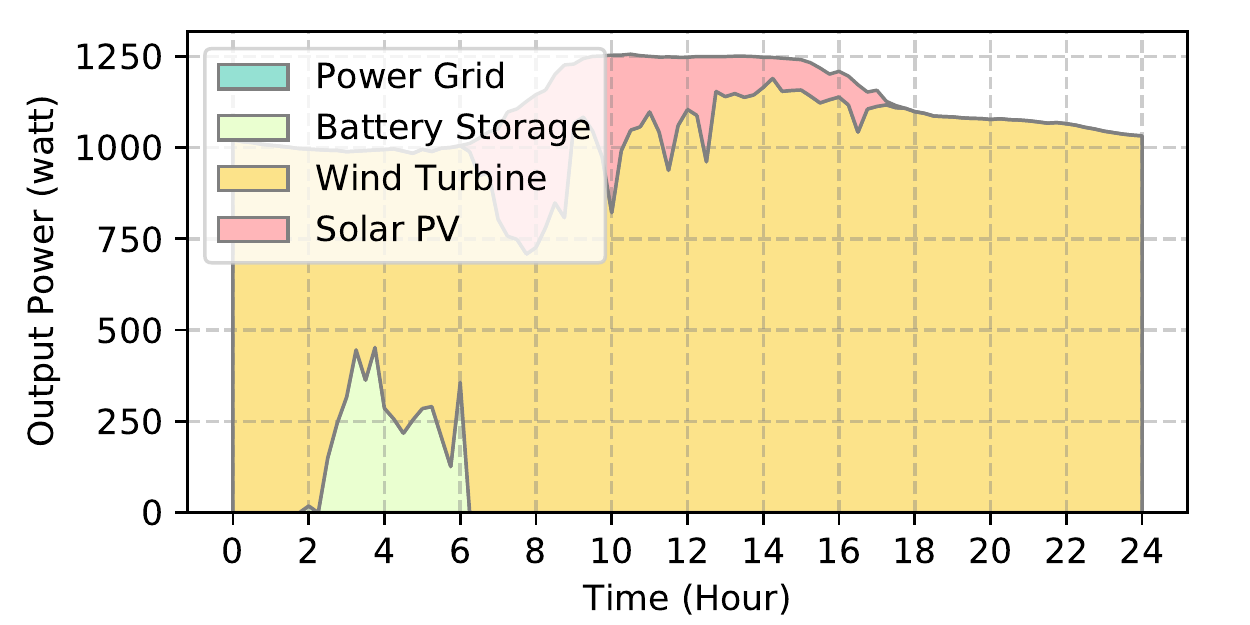}}
 	\hfill
 	\subfigure[The power supply pattern under the cloudy \& middle-wind day]{
 		\includegraphics[width=0.32\textwidth]{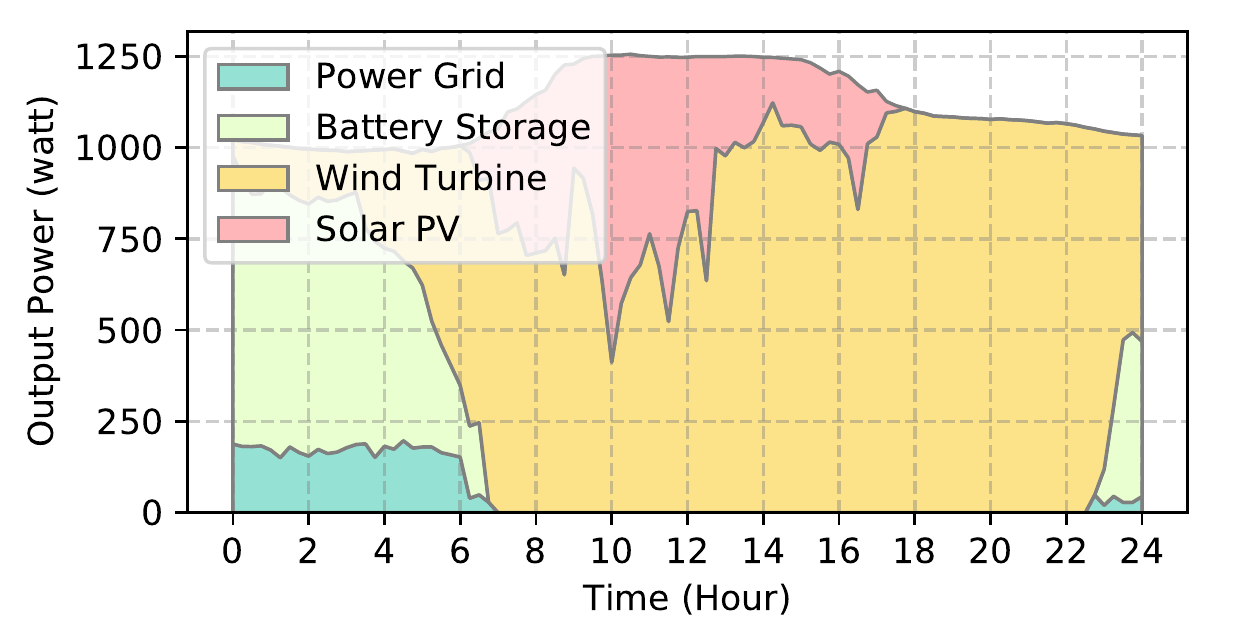}}
 	\hfill
 	\subfigure[The power supply pattern under the cloudy \& low-wind day]{
 		\includegraphics[width=0.32\textwidth]{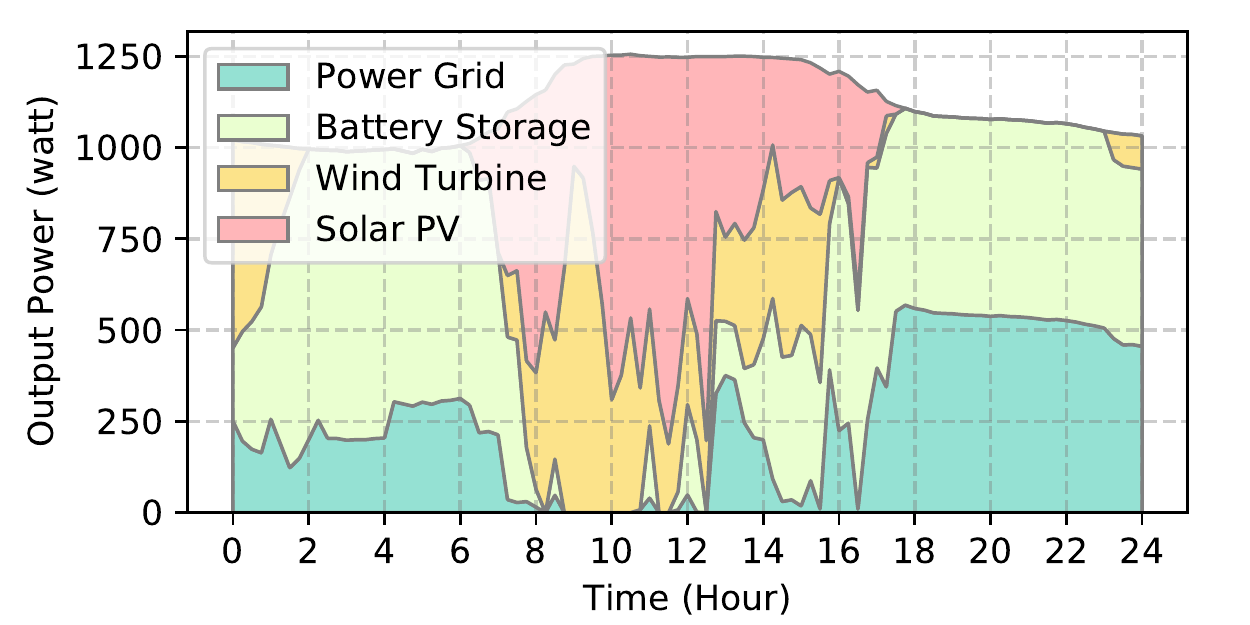}}
 	\caption{The power supply pattern of a single 5G BS at area of office is supplied by different power supply methods under different weather conditions in one day period. }
 	\label{fig:power-pattern-office}
\end{figure*}


\begin{figure*}[!h]
 	 	\subfigure[The power supply pattern under the clear \& high-wind day.]{
 		\includegraphics[width=0.32\textwidth]{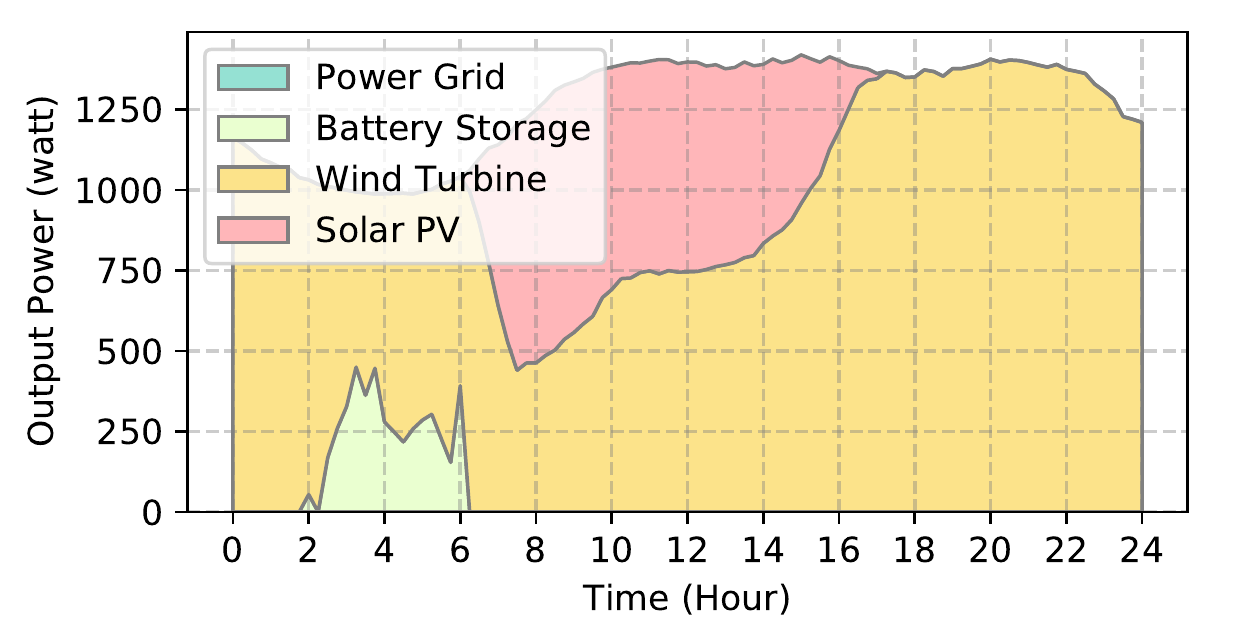}}
 	\hfill
 	\subfigure[The power supply pattern under the clear \& middle-wind day]{
 		\includegraphics[width=0.32\textwidth]{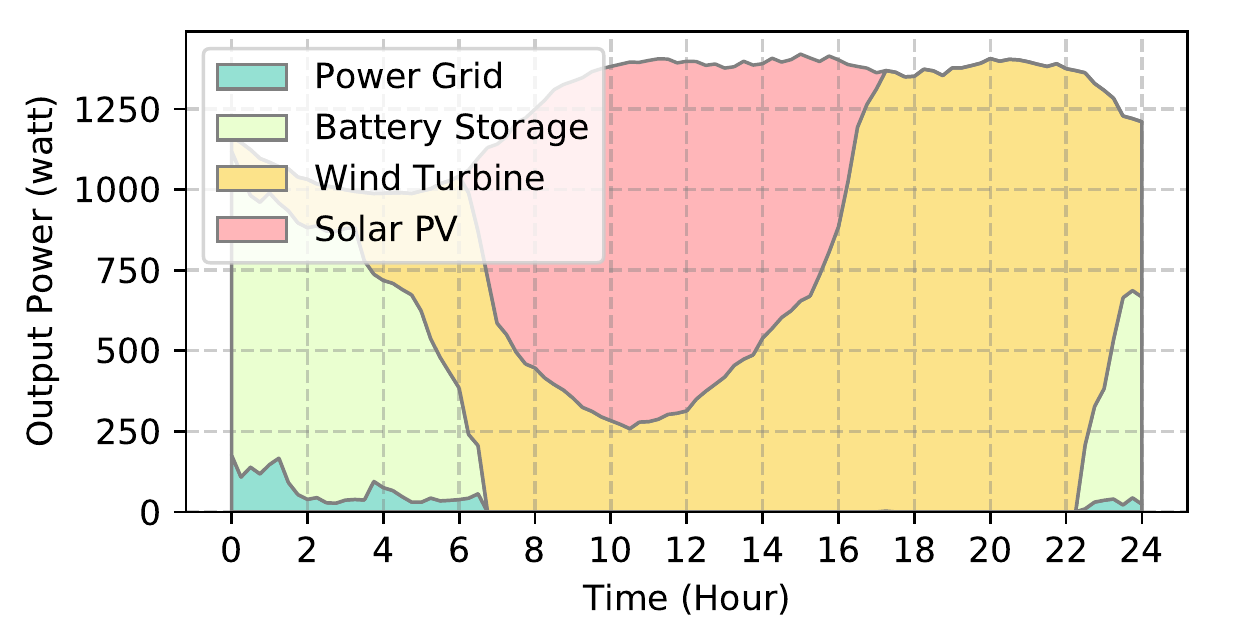}}
 	\hfill
 	\subfigure[The power supply pattern under the clear \& low-wind day]{
 		\includegraphics[width=0.32\textwidth]{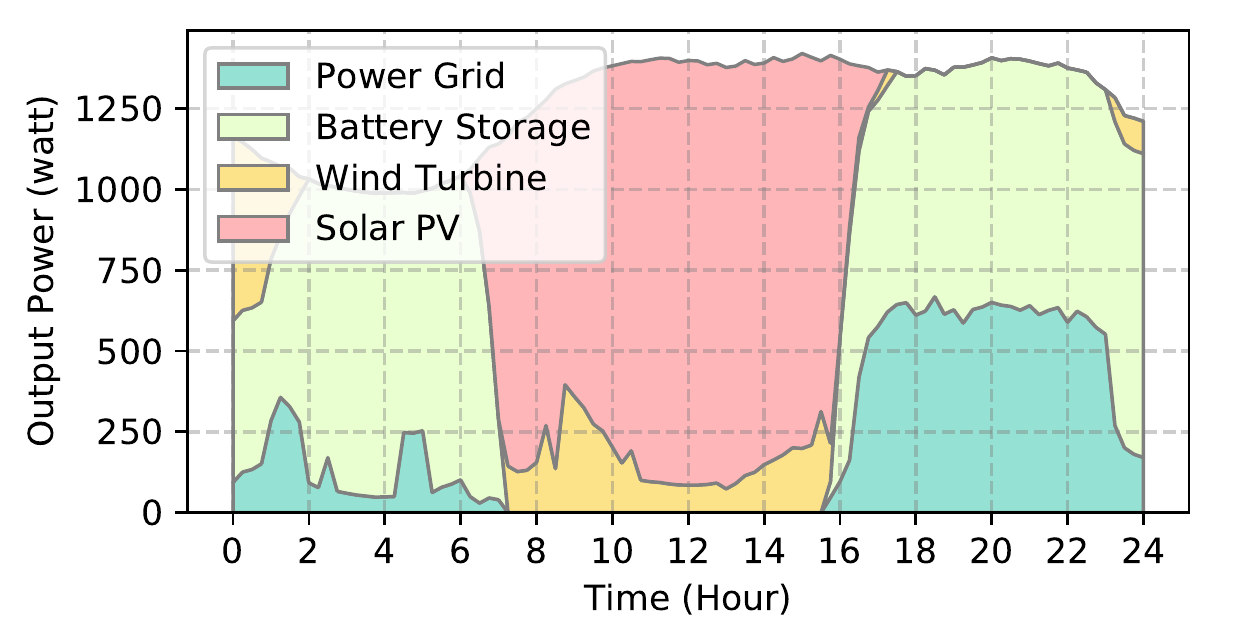}}
 	\\
 	 	\subfigure[The power supply pattern under the partial cloudy \& high-wind day]{
 		\includegraphics[width=0.32\textwidth]{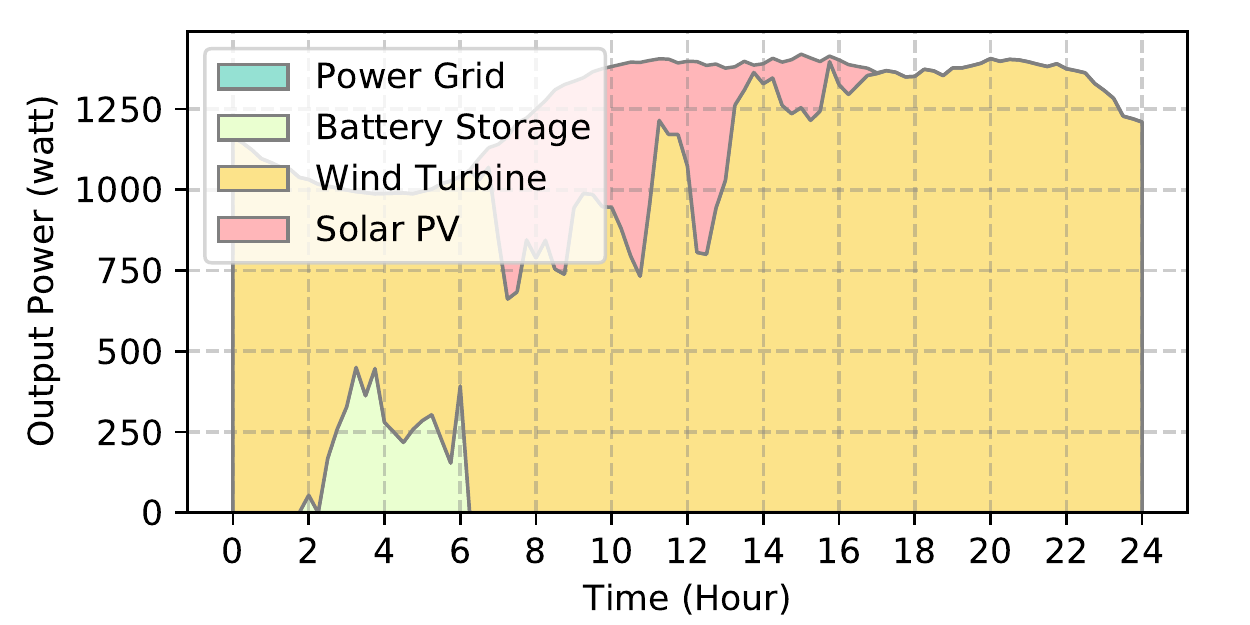}}
 	\hfill
 	\subfigure[The power supply pattern under the partial cloudy \& middle-wind day]{
 		\includegraphics[width=0.32\textwidth]{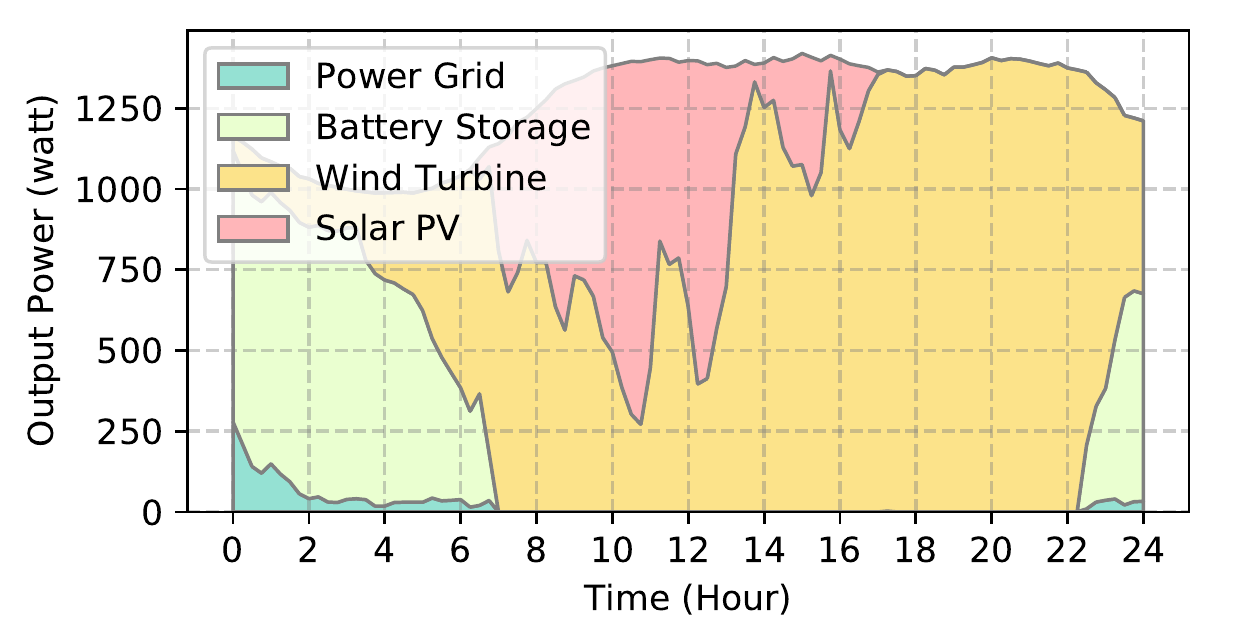}}
 	\hfill
 	\subfigure[The power supply pattern under the partial cloudy \& low-wind day]{
 		\includegraphics[width=0.32\textwidth]{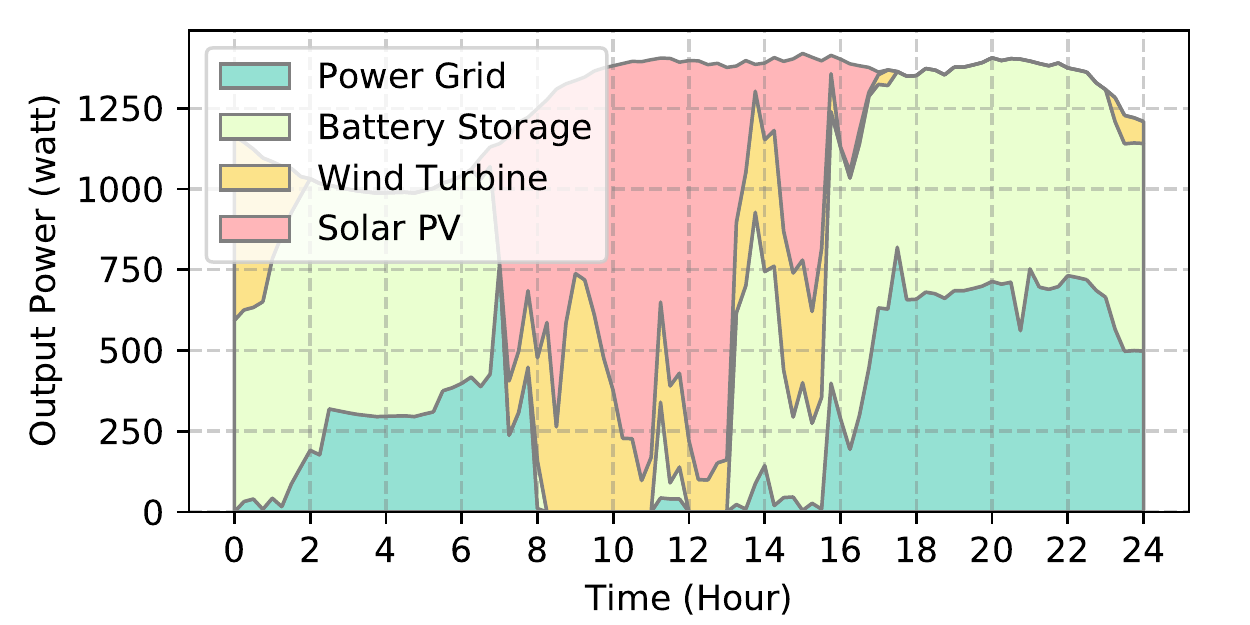}}
 	\\
 	\subfigure[The power supply pattern under the cloudy \& high-wind day]{
 		\includegraphics[width=0.32\textwidth]{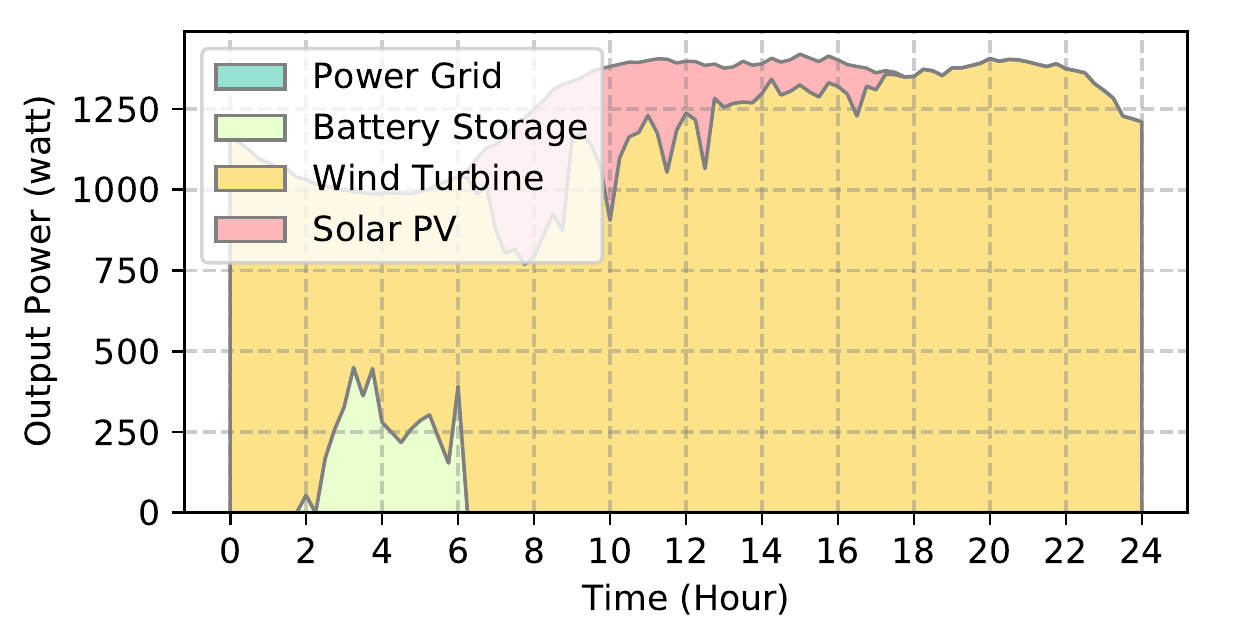}}
 	\hfill
 	\subfigure[The power supply pattern under the cloudy \& middle-wind day]{
 		\includegraphics[width=0.32\textwidth]{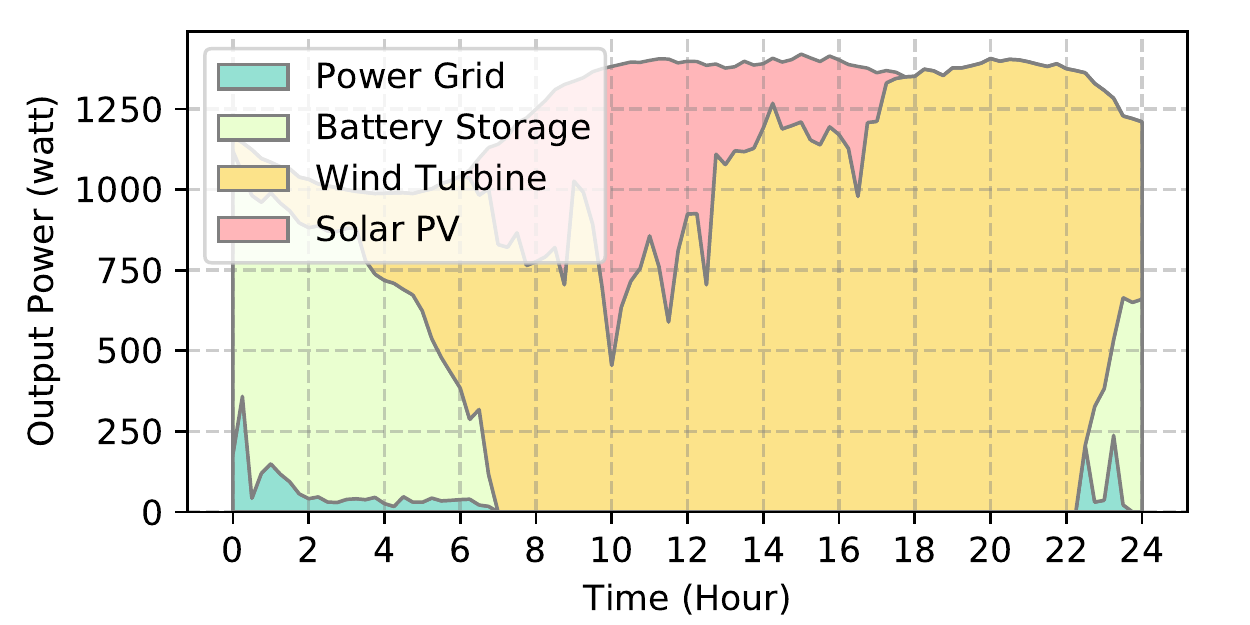}}
 	\hfill
 	\subfigure[The power supply pattern under the cloudy \& low-wind day]{
 		\includegraphics[width=0.32\textwidth]{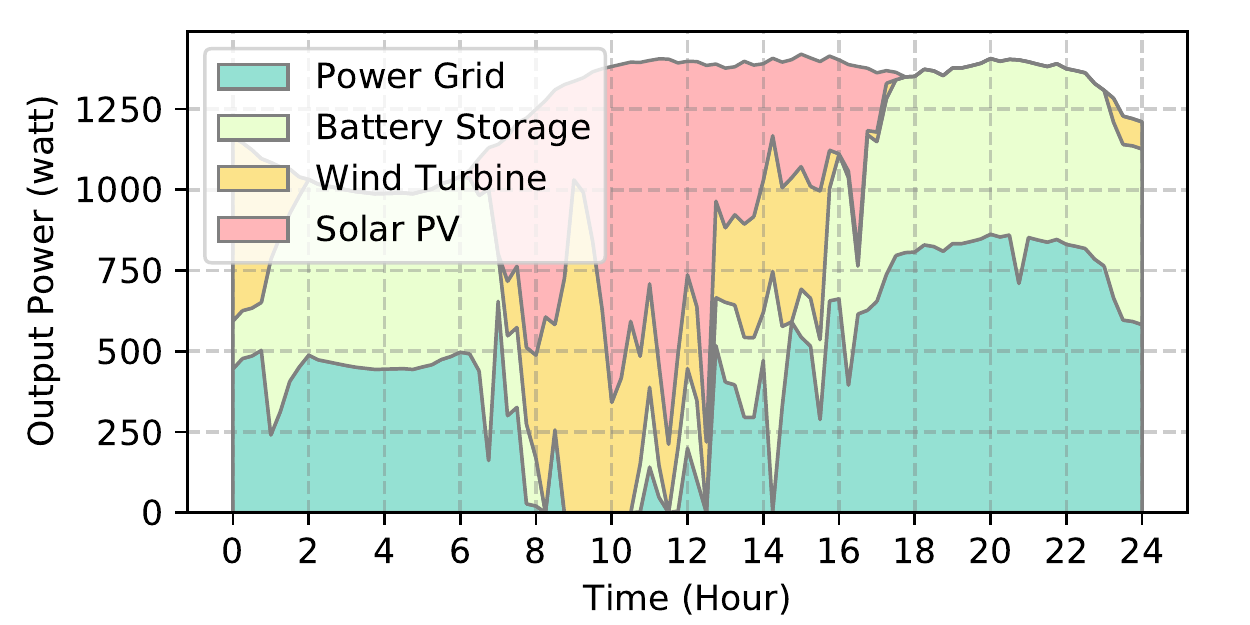}}
 	\caption{The power supply pattern of a single 5G BS at area of comprehensive is supplied by different power supply methods under different weather conditions in one day period. }
 	\label{fig:power-pattern-comprehensive}
\end{figure*}

\end{document}